\documentclass[1p,11pt,sort&compress,plainnat]{elsarticle}
\usepackage{graphicx,bm,enumerate,amssymb,bbm,amsmath}

\newcommand{\jbo}{\mathcal{J}} %bulk current operator
\newcommand{\jb}{J} %bulk current expectation value
\newcommand{\jeo}{j} %edge current operator
\newcommand{\jeoh}{\hat{\jmath}} %edge current operator conserved
\newcommand{\Af}{\mathpzc{A}}
\newcommand{\Afb}{\pmb{\mathpzc{A}}}
\newcommand{\af}{\mathpzc{a}}
\newcommand{\afb}{\pmb{\mathpzc{a}}}

\DeclareFontFamily{OT1}{pzc}{}
\DeclareFontShape{OT1}{pzc}{m}{it}{<-> s * [1.28] pzcmi7t}{}
\DeclareMathAlphabet{\mathpzc}{OT1}{pzc}{m}{it}

\DeclareMathAlphabet{\mathcal}{OMS}{cmsy}{m}{n}

\date{July 25, 2011}

\begin{document}

\author[mit,eth]{Samuel Bieri}
\ead{sbieri@mit.edu}

\author[eth]{J\" urg Fr\" ohlich}
\ead{juerg@itp.phys.ethz.ch}

\address[mit]{Department of Physics, Massachusetts Institute of Technology, Cambridge, Massachusetts 02139, USA}
\address[eth]{Institute for Theoretical Physics, ETH Z\"urich, 8093 Z\"urich, Switzerland}

\title{Effective field theory and tunneling currents in the fractional quantum Hall effect}

\begin{abstract}
We review the construction of a low-energy effective field theory and its state space for ``abelian'' quantum Hall fluids. The scaling limit of the incompressible fluid is described by a Chern-Simons theory in 2+1 dimensions on a manifold with boundary. In such a field theory, gauge invariance implies the presence of anomalous chiral modes localized on the edge of the sample. We assume a simple boundary structure, i.e., the absence of a reconstructed edge. For the bulk, we consider a multiply connected planar geometry. We study tunneling processes between two boundary components of the fluid and calculate the tunneling current to lowest order in perturbation theory as a function of dc bias voltage. Particular attention is paid to the special cases when the edge modes propagate at the same speed, and when they exhibit two significantly distinct propagation speeds.  We distinguish between two ``geometries'' of interference contours corresponding to the (electronic) {\it Fabry-Perot} and {\it Mach-Zehnder} interferometers, respectively. We find that the interference term in the current is absent when exactly one hole in the fluid corresponding to one of the two edge components involved in the tunneling processes lies inside the interference contour (i.e., in the case of a Mach-Zehnder interferometer). We analyze the dependence of the tunneling current on the state of the quantum Hall fluid and on the external magnetic flux through the sample.
\end{abstract}

\begin{keyword}
  fractional quantum Hall effect\sep edge current interferometry\sep dissipationless charge transport\sep Chern-Simons action\sep topological field theory\sep chiral anomaly
\end{keyword}

\maketitle

\section{Introduction}

The theory of the integral and fractional quantum Hall effect (QHE) has quite a long history; see, e.g., \cite{GirvinPrange87}. In \cite{Wen90ED} and in \cite{FrohlichZee91,FrohlichKerler91}, certain topological Chern-Simons theories in 2+1 dimensions have been introduced that are shown to describe the (long-distance and low-energy) scaling limit of the bulk of an incompressible Hall fluid. The proposals in \cite{Wen90ED,FrohlichZee91,FrohlichKerler91} have led to the idea of ``topological order'' and of ``universality classes'' describing the scaling limits of different incompressible Hall fluids; see, e.g., \cite{WenTOQHE95}. More recently, ideas adapted from the theory of the quantum Hall effect have led to a fascinating branch of condensed matter theory that concerns topological order and topological phases of matter exhibiting universal properties in a broader context; \cite{HasanKane10}.

On the experimental side, after the initial discovery of the quantum Hall effect by von Klitzing et al. \cite{Klitzing80}, the next major breakthrough was the discovery of the fractional effect by Tsui, St\"ormer, and Gossard in 1982, \cite{TsuiStormerGossard82}. However, only in recent years, experiments testing mesoscopic aspects of incompressible Hall fluids related, e.g., to the famous edge currents and tunneling properties have become possible; see \cite{JiShtrikman03,CaminoZhouGoldman05, MillerMarcusPfeifferWest07, RaduMarcusPfeifferWest08, BieriSchonenbergerOberholzer09, WilettPfeifferWest10}. Recent experiments probe fractional quantum Hall states by performing transport measurements involving (the tunneling of) edge currents and interference effects.

It appears to be prohibitively difficult to rigorously identify the appropriate universality class describing a quantum Hall state realized in the laboratory from microscopic quantum many-body theory. For this reason, recent experimental activities are crucially important in that they yield evidence for the correct choice of an effective field theory describing a concrete incompressible state corresponding to some plateau of the Hall conductance. In order to interpret experimental data and come up with the correct choice, it is important to work out detailed properties of the topological field theories describing the bulk of incompressible Hall fluids and of the effective field theories \cite{BieriFoehlich11,BoyarskyCheianovFrohlich09,LevkivskyiBoyarskyFrohlichSukhorukov09,LFSprog} describing the degrees of freedom located on the edges of incompressible fluids and, in particular, of the chiral currents circulating along the edges of such fluids.

Thus, the aim of our paper is, on one hand, to present a detailed account of the construction of effective field theories and their state spaces describing abelian incompressible quantum Hall fluids in the scaling limit. This part of the paper is partly a review of results that have been established in previous work; (see \cite{BieriFoehlich11} and refs. given there). A fairly broad description of the underlying physical principles, which we will rely upon in the following, has been given in \cite{BieriFoehlich11}. The present paper extends previous ones by the emphasis it places on the role of gauge invariance and on the structure of superselection sectors encountered in the effective theories of Hall edges in the presence of slowly varying external electromagnetic fields and of essentially static, localized quasi-particles in the bulk of a sample. On the other hand, using these results, we present a general analysis of certain charge transport properties related to the degrees of freedom at the edge of an incompressible Hall fluid. Our explicit formulae may be useful to interpret experimental data. Our results are general enough to be applicable to a description of an arbitrary incompressible Hall fluid supporting quasi-particles with abelian fractional (or braid) statistics. Actually, many of our general ideas easily extend to fluids supporting quasi-particles with non-abelian braid statistics, too.

\subsection{Structure of the paper}

This paper is organized in the following way. In Sect.~\ref{sec:lowenergy}, we review the general construction of an effective field theory in 2+1 dimensions describing the low-energy and long-distance physics of an abelian quantum Hall fluid. In this section, we profit from an introduction to the subject contained in the review paper \cite{BieriFoehlich11}; (readers not familiar with the subject are encouraged to consult that paper or some of the refs. given there).\footnote{Many references to the original literature are omitted in the present
paper but can be found in \cite{BieriFoehlich11}.} We then discuss the constraints on the physical state space of topological field theories describing electron fluids exhibiting the quantum Hall effect.

In Section \ref{sec:GI}, we discuss the boundary terms in the effective action describing an incompressible Hall fluid. The gauge anomaly of the Chern-Simons theory calls for massless (non-topological) field theories in 1+1 dimensions localized on the edge of the sample. We construct vertex operators creating gapless quasi-particles propagating along the edge of an incompressible fluid and we discuss the superselection sectors of the edge theory.

Section \ref{sec:interedgetransport} is devoted to charge transport through tunneling constrictions between disjoint boundary components within the framework of the previous two sections. We construct gauge invariant tunneling operators and calculate explicitly the inter-edge tunneling current to lowest order in perturbation theory. We argue that the appropriate initial state for perturbation theory is a superposition of states in different charge superselection sectors for the two boundary components involved in the transport. This averaging, due to the fractional electric charge of the tunneling quasi-particle, leads to the absence of an interference term in the tunneling current for some geometries of interference contours. We also discuss the effect of time-dependent modifications of the sample caused by external agents on charge transport through the sample. We find that an adiabatic increase of the flux of the external magnetic field through the interference contour does not lead to any measurable signature (``Aharonov-Bohm phase'') in the tunneling current. However, changing the magnetic flux through the interference contour by deforming the edges or dragging bulk quasi-particles out of the contour is expected to have observable consequences and may be used to experimentally probe universal properties of a fractional quantum Hall plateau.

\section{Low-energy description of a quantum Hall fluid}\label{sec:lowenergy}

We consider a quantum Hall fluid confined to a planar domain $\Omega\subset\mathbb{R}^2$ of space and subject to a strong transverse magnetic field. We denote the bulk space-time of the fluid by $\Lambda = \Omega\times\mathbb{R}$. The boundary of space-time is denoted by $\partial\Lambda = \partial\Omega\times\mathbb{R}$ and we let $\mathring\Lambda = \Lambda\setminus\partial\Lambda$ be the interior of $\Lambda$.

From the phenomenology of the incompressible quantum Hall fluid (i.e., exhibiting a positive energy gap above the ground state; see \cite{BieriFoehlich11}, and refs. therein), one derives the following basic equation for the electric current two-form $\jb$ in the 2+1 dimensional bulk $\mathring\Lambda$ of the Hall sample:
\begin{equation}\label{eq:phenomenology}
  \jb(x) = \sigma\, dA(x)\,,\;\; x \in\mathring\Lambda\, .
\end{equation}
Here $\sigma$ is the Hall conductivity in units of $e^2/h$; $d$ denotes an exterior derivative, and $A$ describes the fluctuations of the external electromagnetic gauge potential around the one corresponding to a constant background magnetic field ($B_c = \text{curl} A_c$). For simplicity, we consider a Hall conductivity $\sigma$ that has a constant value throughout the sample $\Omega$. In this paper, we choose units such that $e=h=1$.

From the phenomenological law Eq.~\eqref{eq:phenomenology}, we infer the effective action, $\Gamma[A]$, which is the generating function of electric current correlators in a quantum Hall fluid:
\begin{equation}\label{eq:effaction}
  \Gamma[A] = \frac{\sigma}{2}\int_{\Lambda} A\wedge dA + \Gamma_{\partial\Lambda}[a]\,.
\end{equation}
%Note that there is a factor of 1/2 difference in the definition of the boundary action with respect to the review paper.
The term $\Gamma_{\partial\Lambda}[a]$ is a function of the {\it boundary values} of the gauge field, $a = A|_{\partial\Lambda}$. Up to manifestly gauge-invariant terms, it is uniquely determined by the requirement of gauge invariance of the total action, i.e., $\Gamma[A+d\alpha] = \Gamma[A]$. The boundary term will be discussed in more detail in Sect.~\ref{sec:GI}. The phenomenological law \eqref{eq:phenomenology} is then equivalent to the equation
\begin{equation}
  \jb(x) = \frac{\delta\Gamma[A]}{\delta A(x)}\,,\;\; x\in \mathring\Lambda\,,
\end{equation}
with $\Gamma[A]$ as in \eqref{eq:effaction}.

\subsection{Abelian quantum Hall fluids}

The generating function \eqref{eq:effaction} does not provide complete information about the {\it low-energy effective} degrees of freedom that are involved in a description of a quantum Hall fluid corresponding to a particular plateau of the Hall conductivity $\sigma$. We denote the quantum-mechanical electric current density by $\jbo$. In this paper we discuss ``abelian'' quantum Hall fluids. This means that $\jbo$ is a linear combination of $N$ independent, conserved quasi-particle currents. We denote the quasi-particle current operators by $\jbo_n$. The electric current is written as
\begin{equation}\label{eq:defJ}
  \jbo = \sum_{n=1}^N Q_n \jbo_n\,,
\end{equation}
where the coefficients $Q_n$ are the electromagnetic coupling constants of the quasi-particles carrying the currents $\jbo_n$. In the integral quantum Hall effect, $n$ labels different Landau levels, and all the $Q_n$ can be set to unity. However, for a {\it fractional} fluid, the currents $\jbo_n$ may be carried by ``emergent'' excitations of the electronic system.

The relation of the current operator $\jbo(x)$ to the current density $\jb(x)$ of the previous section is that of an expectation value in a state of the Hall fluid, i.e.,
\begin{equation}
  \jb(x) = \langle \jbo(x) \rangle\,.
\end{equation}
Physically interesting states, $\langle(\cdot)\rangle$, of the system will be discussed later.

The quasi-particle currents in \eqref{eq:defJ} are separately conserved, i.e.,
\begin{equation}\label{eq:conservation}
  d\jbo_n = 0\, .
\end{equation}
This allows us to write the currents in terms of vector potentials $B_n$,\footnote{If the manifold $\Lambda$ has trivial second de Rham cohomology, then closed two-forms on $\Lambda$ are (globally) exact forms. Only in this case, \eqref{eq:Jn} holds for vector potentials $B_n$ that are single-valued on $\Lambda$. Indeed, we have $H^2_{dR}(\Omega\times\mathbb{R})=0$ when $\Omega$ is a disc with a finite number of punctures. We thank Robert Bieri for this comment.}
\begin{equation}\label{eq:Jn}
  \jbo_n = dB_n\, .
\end{equation}
This representation of the currents by potentials $B_n$ gives rise to the gauge symmetry $B_n\mapsto B_n + d\beta_n$. The low-energy action describing the quantum theory of the vector potentials $B_n$ is given by
\begin{equation}\begin{split}\label{eq:cs1}
  S[{\bm B}, A] &= S_{\Lambda}[{\bm B}, A] + S_{\partial\Lambda}\\
  &= \sum_n \int_\Lambda \{ - \frac{\chi_n}{2} B_n \wedge dB_n + Q_n A\wedge dB_n \} + S_{\partial\Lambda}\, .
\end{split}\end{equation}
Here, we use the notation ${\bm B} = (B_n)$. The term $S_{\partial\Lambda}$ depends on the boundary restrictions of the fields. It is partially determined by gauge invariance of the total action $S[{\bm B}, A]$. The boundary term will be discussed in detail in Sect.~\ref{sec:GI}. Below, we will see that $\chi_n = \pm$ is the sign of the Hall conductivity associated with the quasi-particle current $\jbo_n$.

Equation \eqref{eq:cs1} for $S[{\bm B}, A]$ reproduces the effective action $\Gamma[A]$, after integrating out the gauge potentials $B_n$,
\begin{equation}\label{eq:path}
  e^{ 2\pi i\, \Gamma[A] } := \mathcal{Z}^{-1} \int [D{\bm B}]\, e^{ 2\pi i\, S[{\bm B}, A] }\,,
\end{equation}
with $\mathcal{Z} = \int [D{\bm B}]\, e^{ 2\pi i\, S[{\bm B}, 0] }$. Comparing \eqref{eq:path} with \eqref{eq:effaction}, one finds the following relation between the parameters $Q_n$ and $\chi_n$ in the low-energy action \eqref{eq:cs1}, and the Hall conductivity, $\sigma$,
\begin{equation}\label{eq:qconstraint}
  \sigma = \sum_n \chi_n Q_n^2 := {\bm Q}\cdot {\bm Q}\, .
\end{equation}
In the following, we use the convention that $\chi_{n\leq N_e}=1$ and $\chi_{n> N_e}=-1$, and define
\begin{equation}
  \sigma_e = \sum_{n=1}^{N_e} Q_n^2\,,\,\,\sigma_h = \sum_{n=N_e+1}^{N} Q_n^2\,,
\end{equation}
so that
\begin{equation}
  \sigma = \sigma_e - \sigma_h\, .
\end{equation}
We see that the parameter $\chi_n$ determines the sign of the Hall conductivity corresponding to the current $\jbo_n$ (if $Q_n\neq 0$). Therefore, the channels with $\chi_n=-1$ and $\chi_n=1$ may be called {\it hole} and {\it electron channels}, respectively.

Equation \eqref{eq:cs1} defines the action of a topological field theory (TFT) for $N$ abelian gauge fields $B_n$ in 2+1 dimensions. In the next section, we discuss the state space of such a TFT. For a discussion of states for more general TFTs describing quantum Hall fluids with non-abelian braid statistics, see \cite{BieriFoehlich11} and references therein.

\subsection{\label{sec:bulkstates}Bulk states of an abelian quantum Hall fluid}

The {\it observables} of the field theory \eqref{eq:cs1} are gauge-invariant and ``topological'' (i.e. metric-independent) operators. They are ``Wilson-loops'' given by the expressions
\begin{equation}\label{eq:wilson}
  W[{\bm B}, {\bm A}] = \exp \{ 2\pi i \sum_n \int_\Lambda B_n \wedge dA_n \}\, .
\end{equation}
The observables $W[{\bm B}, {\bm A}]$ are functionals of some conserved current distribution,
\begin{equation}
  (\bar \jb_n)_{n=1}^N :=  \bar {\bm \jb} = d{\bm A} %\, , n=1\ldots N\,,
\end{equation}
and of the gauge potentials $B_n$. We require the currents $(\bar J_n)$ to vanish at the boundary $\partial\Lambda$ of space-time, i.e., $dA_n|_{\partial\Lambda}=0$.

In order to discuss the space of states of the (bulk) Chern-Simons theory, we choose a time slice, e.g. $\{ t = 0 \}$, of the space-time manifold $\Lambda$. A sector of states is then indexed by the current distribution $\bar \jb_n = dA_n$ on $\Omega$ at $t=0$,
\begin{equation}\label{eq:statespace}
  [ \bar {\bm \jb}(t=0) ]\, .
\end{equation}
In the abelian theories considered in this paper, the sectors $[ \bar {\bm \jb}(t=0) ]$ are {\it one-dimensional}, and the phase of a state is given by the {\it history} of the current distribution, $\bar \jb_n(t<0)$. The path integral induces a natural inner product on the state space of the theory. Sectors with different current distributions at $t=0$ are orthogonal. We pick two states in a sector,
\begin{equation}\label{eq:csstates}
  | \bar {\bm \jb}^{(1)} \rangle, | \bar {\bm \jb}^{(2)} \rangle \in [ \bar {\bm \jb}(t=0) ]\, .
\end{equation}
The inner product of these states is defined as the expectation value of the corresponding Wilson loop operator, Eq.~\eqref{eq:wilson},
\begin{equation}\label{eq:transition}
  \langle \bar {\bm \jb}^{(2)} | \bar {\bm \jb}^{(1)} \rangle = \langle W[{\bm B}, {\bm A}] \rangle = \mathcal{Z}^{-1} \int [D{\bm B}]\, W[{\bm B}, {\bm A}]\, e^{2\pi i\, S[{\bm B}, A]}\, ,
\end{equation}
with $dA_n = \bar \jb_n$ such that $\bar\jb_n(t<0) = \bar\jb^{(1)}_n(t)$ and $\bar\jb_n(t>0) = -\bar\jb^{(2)}_n(-t)$.\footnote{Note that the {\it temporal component} of %$*\bar\jb_n(t)$
the current (i.e., the charge density) is continuous at $t=0$; i.e., let $*\bar\jb_n = \bar\jb_{n}^\mu dx_\mu$, where $*$ is the Hodge dual; then $\bar\jb_{n}^0(t=0^+)= \bar\jb_{n}^0(t=0^-)$. The spatial components of the current, however, have a kink at $t=0$ for non-static charge distributions, i.e., $\bar\jb_{n}^i(t=0^+) = -\bar\jb_{n}^i(t=0^-)$.}

We consider a quantum Hall fluid with some given static quasi-particle charge distribution. Then, we may include the Wilson operators defining the charge distribution into the action. That is, in the presence of a quasi-particle background, the bulk action can be written as
\begin{equation}\label{eq:cs2}
  S_{\Lambda}[{\bm B}, {\bm Q} A + {\bm A}] = \sum_n \int_\Lambda \{-\frac{\chi_n }{2} B_n \wedge dB_n + (Q_n A + A_n)\wedge dB_n \}\, .
\end{equation}
From \eqref{eq:defJ} and \eqref{eq:Jn}, the electric current in the bulk acquires an additional contribution due to the insertion of the Wilson loop operator in the expectation value,
\begin{equation}\label{eq:bulkcurrent}
  \jb = \langle\jbo\rangle = \sigma dA + \sum_n \chi_n Q_n\, dA_n\, .
\end{equation}

In order to simplify our notation, we combine the gauge potentials $A$ and $A_n$ into a single field, $\Af_n$,
\begin{equation}
  \Af_n = Q_n A + A_n\, .
\end{equation}
The quasi-particle currents are now given by
\begin{equation}\label{eq:bulkqpcurrent}
  \langle \jbo_n \rangle = \chi_n d\Af_n = \chi_n Q_n dA + \chi_n dA_n\, . % = \chi_n d\Af}_n = \chi_n Q_n dA + \chi_n \bar \jb_n\, .
\end{equation}

\subsubsection{Monodromy and braiding of localized bulk states}

For simplicity, we assume in the following that the electromagnetic field fluctuations vanish, i.e., $dA=0$. Let ${\bm q} = (q_n) \in \mathbb{R}^N$ and let
\begin{equation}\label{eq:locstate}
  |{\bm q}, z\rangle \in [{\bm q},z]
\end{equation}
be a state with localized charge such that
\begin{equation}\label{eq:bulkcharge}
  (\int_\mathcal{O} d^2x\, \mathcal{J}^0_n)\, |\bm{q}, z \rangle = \chi_n q_n\, |\bm{q}, z\rangle\,,
\end{equation}
where $z\in\mathcal{O}\subset\Omega$; $\mathcal{O}$ is a small region in space containing the point $z$. The state $|{\bm q}, z\rangle$ corresponds to the insertion of a Wilson loop operator \eqref{eq:wilson} with a localized source for the field $A_n$, i.e., $\int_{\partial\mathcal{O}} A_n = q_n$. The {\it electric charge} of this state located in the region $\mathcal{O}$ is given by [see Eq.~\eqref{eq:defJ}]
\begin{equation}\label{eq:chargeel}
  Q_{el}({\bm q}) = \sum_n \chi_n Q_n q_n := {\bm Q}\cdot {\bm q}\, .
\end{equation}
The {\it (conformal) spin}, $s_{\bm q}$, of the state \eqref{eq:locstate} is determined by considering the operation of a full rotation around the origin, $U_{rot}(2\pi)$. It can be shown that (see, e.g., \cite{FrohlichMarchetti89,FrohlichKerler91,FrohlichThiran94})
\begin{equation}
  U_{rot}(2\pi) | {\bm q}, z \rangle = \exp ( 2\pi i s_{\bm q} ) | {\bm q}, z \rangle \,,
\end{equation}
where
\begin{equation}\label{eq:qspin}
  s_{\bm q} = \frac{{\bm q}\cdot {\bm q}}{2}\, .
\end{equation}

\begin{figure}[!h]
\center
\includegraphics[width=0.6\textwidth]{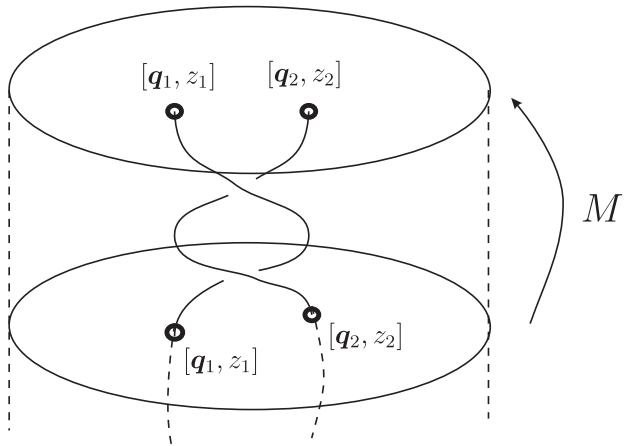}
\caption{Monodromy operation of a state with two sources located at $z_1,z_2\in \Omega$.
\label{fig:monodromy}
}
\end{figure}

Next, let us consider the monodromy of a state in a sector with two sources located at $z_1$ and $z_2$,
\begin{equation}
  [{\bm q}_1,z_1] \otimes [{\bm q}_2,z_2]\,,
\end{equation}
under a twist through an angle of $2\pi$ (see Fig.~\ref{fig:monodromy}). The monodromy matrix is defined by
\begin{equation}
  M({\bm q}_1,{\bm q}_2) = \exp \{-2\pi i (s_{\bm q_1} + s_{\bm q_2}) \} U_{rot}(2\pi)|_{[{\bm q_1},z_1]\otimes[{\bm q_2},z_2]}\,,
\end{equation}
i.e., it represents the action of taking one source around the other source without rotating the spins. For $z_1\approx z_2$, we have that
\begin{equation}
  [{\bm q}_1,z_1]\otimes[{\bm q}_2,z_2] \simeq [{\bm q}_1+{\bm q}_2,z_1]\,,
\end{equation}
and, using \eqref{eq:qspin}, one immediately finds that
\begin{equation}\label{eq:monodromy}
  M({\bm q}_1,{\bm q}_2) = \exp\{ 2\pi i\, {\bm q}_1 \cdot {\bm q}_2 \}\, .
\end{equation}

A {\it charge-statistics connection} holds for the states of the Chern-Simons theory \cite{FrohlichMarchetti89}: For two sources with identical charge vector ${\bm q}$, one defines a ``half-monodromy'' by
\begin{equation}
  M_{1/2}({\bm q}) = \exp \{-2\pi i s_{\bm q} \} U_{rot}(\pi)|_{[{\bm q},z]\otimes[{\bm q},-z]}\,,
\end{equation}
where $U_{rot}(\pi)$ is a rotation through an angle $\pi$ around the origin. This operation represents a {\it permutation} of the two sources. One can show that
\begin{equation}\label{eq:chargestatistics}
  M_{1/2}({\bm q}) = \exp\{2\pi i s_{\bm q}\}\,,
\end{equation}
which, in view of \eqref{eq:qspin}, provides a relation between charge and statistics of localized sources.\footnote{One may also call \eqref{eq:chargestatistics} a {\it conformal spin-statistics connection}. Notice, however, that we are considering fully polarized quantum Hall fluids where the physical electron spin does not play a role.}

It is known that the quantum states of topological field theories in 2+1 dimensions, here \eqref{eq:locstate}, are in one-to-one correspondence with conformal blocks ($n$-point functions) of conformal field theories (CFTs) in two dimensions. This fact was conjectured by Witten in \cite{Witten89} and elaborated and extended in \cite{FrohlichKing89} and many other papers. See also \cite{FrohlichPedriniSchweigertWalcher01} for applications to the QHE. The monodromy and braiding of states exposed in this section [Eqs. \eqref{eq:qspin}, \eqref{eq:monodromy}, \eqref{eq:chargestatistics}] may also be obtained from the corresponding CFT on $\Omega$. Here, we do not use this point of view. However, our construction of massless field theories localized on the space-time boundary $\partial\Lambda$ in Sect.~\ref{sec:GI} is another manifestation of this correspondence.

\subsubsection{Classification of quantum Hall fluids}

In this section, we review the constraints on the state space of the abelian Chern-Simons theory imposed by the physical requirement that it describes an incompressible quantum Hall fluid. On physical grounds, we start from the following two postulates:
\begin{enumerate}[A)]\label{eq:physical}
  \item Among the set of states, there exist (localized) states corresponding to single electrons or holes, i.e., carrying {\it electric charge $\pm 1$}.\footnote{The charge of an electron $e=1$, in our units.} Multi-electron and -hole states are supposed to obey {\it Fermi statistics}.
  \item An arbitrary state must be single-valued in the positions of electrons and holes.
\end{enumerate}

Using the results presented in the last section, we are prepared to discuss the implications of postulates A) and B) on the structure of the state space. For simplicity, we assume that electron or hole states are not composed of sources with different chiralities $\chi_n$. Thus, we can discuss the electron- ($\chi_n=1$) and hole- ($\chi_n=-1$) channels separately. Let $N = N_e + N_h$, where $N_e$ and $N_h$ are the number of electron and hole channels, respectively. In the following, we restrict our attention to electron channels.

Postulate~A), concerning the existence of localized states corresponding to electrons, with charge vector ${\bm q}_{e}\in \mathbb{R}^{N_e}$, implies that
\begin{equation}\label{eq:chargecond}
  {\bm Q}_e \cdot {\bm q}_{e} = 1
\end{equation}
for the electric charge by Eq.~\eqref{eq:chargeel}, and
\begin{equation}\label{eq:spincond}
  {\bm q}_{e} \cdot {\bm q}_{e} \in 2 \mathbb{Z} + 1\,,
\end{equation}
reflecting Fermi statistics by \eqref{eq:chargestatistics}. The solutions ${\bm q}_e$ to Eq.~\eqref{eq:spincond} span an $N_e$-dimensional, odd-integral lattice $\Gamma_e\subset\mathbb{R}^{N_e}$. From \eqref{eq:chargecond}, it follows that ${\bm Q}_e = (Q_n)_{n=1}^{N_e}$ must be a ``visible vector'' in the lattice dual to $\Gamma_e$, i.e., ${\bm Q}_e \in \Gamma_e^{*}$. For more details on integral lattices, see \cite{FrohlichThiran94}.

Postulate B) implies that the monodromy of a state with a localized source ${\bm q}_e$ corresponding to an electron and any other physical localized source ${\bm q}$ must be the identity. From Eq.~\eqref{eq:monodromy}, one finds that
\begin{equation}\label{eq:monodromystates}
  {\bm q}\cdot {\bm q}_e \in \mathbb{Z}\, .
\end{equation}
It follows from \eqref{eq:monodromystates} that charge vectors ${\bm q}$ of physical localized sources lie in the lattice $\Gamma_e^{*}$ dual to the lattice $\Gamma_e$ of electron states,
\begin{equation}
  {\bm q} \in \Gamma^{*}_e\, .
\end{equation}
Note that $\Gamma_e\subset\Gamma^*_e$, for integral lattices.

The same arguments can be repeated for hole channels. The result is an $N_h$-dimensional, odd-integral lattice $\Gamma_h\subset\mathbb{R}^{N_h}$ of ``hole'' states and ${\bm Q}_h = (Q_n)_{n=N_e+1}^{N} \in \Gamma_h^*$ is a visible vector in the dual lattice $\Gamma_h^*$. The full space of physical states is labeled by vertices in $\Gamma_e^*\oplus\Gamma_h^*$.

These results can now be combined with Eq.~\eqref{eq:qconstraint},
\begin{equation}
  \sigma = {\bm Q}\cdot{\bm Q} = {\bm Q}_e\cdot{\bm Q}_e - {\bm Q}_h\cdot{\bm Q}_h = \sigma_e - \sigma_h\, .
\end{equation}
From the fact that $\Gamma_{e}$ and $\Gamma_{h}$ are integral lattices and ${\bm Q}_{e,h}\cdot{\bm Q}_{e,h}\in \Gamma_{e,h}^*$, it follows that $\sigma_{e}$ and $\sigma_h$ take rational values. This implies that the Hall conductivity $\sigma$ is rational; (see also \cite{FrohlichStuderThiran95,FrohlichKerlerStuderThiran95}).

\section{Gauge invariance and the edge action}\label{sec:GI}

\begin{figure}[!h]
\center
\includegraphics[width=0.8\textwidth]{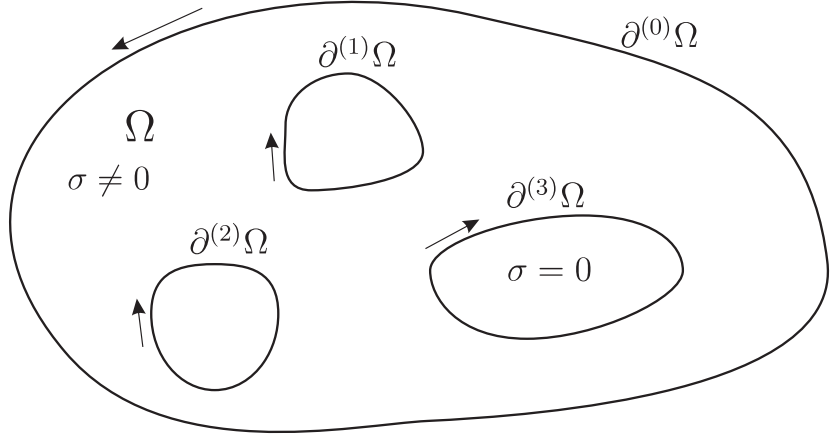}
\caption{The incompressible quantum Hall fluid ($\sigma\neq0$) is located on a connected two-dimensional planar domain of space. The orientation of the different boundary components is indicated by arrows.
\label{fig:sample}
}
\end{figure}

In this section we discuss the boundary terms in the total action, \eqref{eq:cs1}, describing the scaling limit of the incompressible quantum Hall fluid. These boundary terms have to be present because of the requirement of gauge invariance of the total action. In general, the incompressible quantum Hall fluid is located on a planar domain $\Omega\subset \mathbb{R}^2$ of space, as depicted in Fig.~\ref{fig:sample}. The boundary of the fluid can be decomposed into a disjoint union of connected components, $\partial\Omega = \cup_k \partial^{(k)}\!\Omega$.
For most parts of the discussion in this section, it is sufficient to consider a single edge component, i.e., we can restrict our attention to a fluid on a disc with connected boundary $\partial\Omega$. The generalization to a disconnected boundary is straightforward and will, in most cases, not be discussed explicitly.

In reality, the boundary structure of a quantum Hall fluid may be quite complicated, which goes under the notion of {\it edge reconstruction}, see e.g. \cite{Yang03}. We do not consider edge reconstruction in the present paper. Instead, we assume that the edge $\partial\Omega$ of the quantum Hall fluid is sharp and has ``minimal'' internal structure. As we will see, the degrees of freedom localized on the edge are then strongly constrained by the bulk Chern-Simons action. A ``{\it holographic principle}'' between bulk and edge properties becomes manifest.

We start by considering the local symmetries of the bulk action \eqref{eq:cs2}. Gauge transformations of the fields $B_n$ and of the external potentials $\Af_n$ are given by
\begin{equation}\label{eq:gtb}
  B_n \mapsto B_n + d\beta_n\,,
\end{equation}
and
\begin{equation}\label{eq:gta}
\begin{split}
%  A &\mapsto A + d\alpha\,,\\
  \Af_n &\mapsto \Af_n + d \alpha_n\,,
\end{split}
\end{equation}
where $\beta_n$ and $\alpha_n$ are scalar (gauge) functions on $\Lambda$. For gauge functions vanishing at the boundary $\partial\Lambda$, the action \eqref{eq:cs2} is invariant under \eqref{eq:gtb} and \eqref{eq:gta}. However, for gauge functions not vanishing at the boundary, one finds the anomaly
\begin{equation}\label{eq:gtG}
  S_{\Lambda}[{\bm B} + d{\bm \beta}, \Afb + d{\bm \alpha}] = S_{\Lambda}[{\bm B}, \Afb] + \frac{1}{2} \sum_n \int_{\partial\Lambda} ( - \chi_n \beta_n + 2 \alpha_n )\, db_n\,,
\end{equation}
where $b_n = B_n|_{\partial\Lambda}$ are the components of the field $B_n$ tangential to the boundary $\partial\Lambda$.

The gauge-dependent action \eqref{eq:cs2} cannot be used in a path integral formulation of the theory. For a consistent description of the system, the gauge anomaly \eqref{eq:gtG} must be canceled by the anomalies of additional terms in the total action. We define
\begin{equation}\label{eq:Stot}
  S[{\bm B}, \Afb] = S_{\Lambda}[{\bm B}, \Afb] + S_{\partial\Lambda}[{\bm b}, {\bm \phi}, \Afb]\, ,
\end{equation}
and choose $S_{\partial\Lambda}[{\bm b}, {\bm \phi}, \afb]$ such that the total action, $S[{\bm B}, \Afb]$, is invariant under the transformations \eqref{eq:gtb} and \eqref{eq:gta}. Similarly to $b_n$, the fields $\af_n$ are the boundary values of the tangential components of the external gauge fields, $\af_n = \Af_n|_{\partial\Lambda}$. The scalar fields $\phi_n$ in the boundary action, $S_{\partial\Lambda}[{\bm b}, {\bm \phi}, \afb]$, correspond to the gauge-freedom in the choice of the potentials $b_n$ \cite{FrohlichKerler91,FrohlichStuder93,MooreSeiberg89}. Under the gauge transformation \eqref{eq:gtb}, they transform as
\begin{equation}
  \phi_n \mapsto \phi_n - \beta_n|_{\partial\Lambda}\, .
\end{equation}

A boundary action that cancels the gauge anomaly in \eqref{eq:gtG} is given by
\begin{equation}\label{eq:edgeaction}\begin{split}
  S_{\partial\Lambda}[{\bm b}, {\bm \phi}, \afb] = \frac{1}{2} \sum_n \int_{\partial\Lambda} \sqrt{|g_n|} d^2\xi
  &\{ \frac{1}{2} ( \partial_\mu\phi_n + b_{n\mu} ) ( \partial_\nu\phi_n + b_{n\nu})\, g_n^{\mu\nu}\\ &+ ( 2\af_{n\mu} - \chi_n b_{n\mu} ) \epsilon^{\mu\nu} (\partial_\nu \phi_n + b_{n\nu}) \}\, .
\end{split}\end{equation}
Note that the first term in \eqref{eq:edgeaction}, proportional to $g^{\mu\nu}_n$, is a gauge-invariant quantity. Hence the matrices $(g^{\mu\nu}_n)$ are {\it not} determined by the requirement of anomaly cancelation. This reflects the fact that the topological field theory describing the bulk $\mathring\Lambda$ only enforces the gauge symmetries and the values of the parameters \{$N$, $N_e$, $(Q_n)$ \} of the boundary action. However, the bulk theory does not specify the dynamical properties of the edge degrees of freedom.

For simplicity, let us choose a metric $g_n^{\mu\nu}$ in \eqref{eq:edgeaction} of the form
\begin{equation}\label{eq:metricchoice}
  (g^{\mu\nu}_n) = diag(u_n^{-1},-u_n)\,,
\end{equation}
where $u_n>0$ are the {\it propagation speeds} of the edge modes $\phi_n$. In the boundary action \eqref{eq:edgeaction}, $d^2\xi = d\xi\wedge dt$, where $t$ is time and $\xi$ is an appropriate spatial coordinate on the boundary $\partial\Omega$.\footnote{The coordinate $\xi$ is chosen to increase in the direction of {\it positive orientation} of $\partial\Omega$. Note that $\Omega$ is planar and the orientation of $\partial\Omega$ can be consistently chosen, even in the case of a disconnected boundary (see Fig.~\ref{fig:sample}).} With the choice \eqref{eq:metricchoice}, we have $g_n = \det (g_n^{\mu\nu}) = -1$. Note that, in the total action \eqref{eq:Stot}, we have only kept gauge-invariant terms that are quadratic in fields and derivatives. Possible higher-order terms (e.g, a Maxwell term) are {\it irrelevant} (in renormalization group jargon) in the infrared description we are considering here.

\subsection{Effective action describing the boundary}

It is convenient to integrate out the bulk gauge potentials $B_n$ and to introduce an effective action localized on the boundary, $S_{\partial\Lambda}[{\bm \phi}, \afb]$, in such a way that the following equation holds:
\begin{equation}\label{eq:defSboundary}
  \mathcal{Z}^{-1} \int [D{\bm B}]\, e^{2\pi i\, S[{\bm B}, \Afb]} =  e^{2\pi i\, \Gamma_{\Lambda}[\Afb]}\; \mathcal{\tilde Z}^{-1} \int [D{\bm \phi}]\, e^{2\pi i\, S_{\partial\Lambda}[{\bm \phi}, \afb] }\,,
\end{equation}
with $\mathcal{\tilde Z} = \int [D{\bm \phi}]\, e^{2\pi i\, S_{\partial\Lambda}[{\bm \phi}, 0] }$. Integration over $B_n$ is easy, since the action is quadratic. To do so, one can use the equations of motion to eliminate $B_n$ from the action. The field equations derived from the total action, Eq.~\eqref{eq:Stot}, using \eqref{eq:cs2} and \eqref{eq:edgeaction}, are
\begin{equation}\label{eq:motionbulk}
  \chi_n dB_n = d\Af_n
\end{equation}
on $\mathring\Lambda$, and
\begin{subequations}\label{eq:motionedge}
\begin{align}
  (g^{\mu\nu}_n - \chi_n \epsilon^{\mu\nu})(\partial_\nu\phi_n + b_{n\nu}) &= 0\,,\label{eq:motionedge1}\\
  g^{\mu\nu}_n \partial_\mu (\partial_\nu\phi_n + b_{n\nu}) &= 2  d\af_n - \chi_n db_n\,, \label{eq:motionedge2}
\end{align}\end{subequations}
on the boundary $\partial\Lambda$. Equations~\eqref{eq:motionbulk} and \eqref{eq:motionedge1} result from the variation of the action with respect to the gauge potential $B_n$ and Eq.~\eqref{eq:motionedge2} follows from the variation with respect to the field $\phi_n$.

Equation \eqref{eq:motionbulk} is solved by
\begin{equation}\label{eq:solveB}
  \chi_n B_{n\mu} = \Af_{n\mu} + \partial_\mu F_n\,,
\end{equation}
where $F_n$ are arbitrary real single-valued functions on $\Lambda$. It is convenient to shift the integration variables for the remaining path integral over the boundary fields $\phi_n$ to new fields $\tilde \phi_n$,
\begin{equation}\label{eq:defphitilde}
  \chi_n \phi_n = \tilde \phi_n - F_n|_{\partial\Lambda}\, .
\end{equation}
Finally, for the effective action, one finds
\begin{equation}\label{eq:effactionbulk}
  \Gamma_{\Lambda}[\Afb] = \frac{1}{2}\sum_n \chi_n \int_{\Lambda} \Af_n \wedge d \Af_n
\end{equation}
in the bulk, and
\begin{equation}\label{eq:effactionedge}
  S_{\partial\Lambda}[\tilde {\bm \phi}, \afb] = \frac{1}{2}\sum_n \int_{\partial\Lambda} d^2\xi\, \{ \frac{1}{2} ( \partial_\mu\tilde\phi_n + \af_{n\mu}) ( \partial_\nu\tilde\phi_n + \af_{n\nu})\, g_n^{\mu\nu} + \chi_n\, \af_{n\mu} \epsilon^{\mu\nu} \partial_\nu \tilde \phi_n \}
\end{equation}
on the boundary. From \eqref{eq:solveB} and \eqref{eq:defphitilde} it follows that the new boundary modes $\tilde \phi_n$ transform as
\begin{equation}\label{eq:phitildetransf}
  \tilde \phi_n \mapsto \tilde \phi_n - \alpha_n|_{\partial\Lambda}
\end{equation}
under the remaining gauge transformations \eqref{eq:gta}.

Note that Eq.~\eqref{eq:motionedge1} has {\it not} been used in the effective action \eqref{eq:effactionedge}. It plays the role of {\it chiral constraints} on the fields $\tilde\phi_n$. Taking into account \eqref{eq:motionedge1} at this stage would make it impossible to work in a standard path integral formalism. Eq.~\eqref{eq:effactionedge} is the action of free massless Bose fields $\tilde\phi_n$ (``edge modes'') coupled to the external gauge potentials $\af_{n\mu}$. However, only one chiral component of $\tilde\phi_n$ couples to the gauge field; the electromagnetic coupling of the other component of $\tilde\phi_n$ vanishes.

In the following, we exclusively work with the edge action $S_{\partial\Lambda}[\tilde {\bm \phi},\afb]$, Eq.~\eqref{eq:effactionedge}. We omit the tilde on the new boundary modes $\tilde{ \phi}_n$ and write ${ \phi}_n$ instead, remembering that they transform like \eqref{eq:phitildetransf} under gauge transformations.

The boundary term $\Gamma_{\partial\Lambda}$ in the generating function of current correlators, Eq.~\eqref{eq:effaction}, may be obtained by performing the integration over the fields $\phi_n$ in \eqref{eq:defSboundary}. However, here we do not need the explicit expression for $\Gamma_{\partial\Lambda}$, so this calculation is left as an exercise to the reader.

\subsection{Chiral edge currents}

The bulk electric current was given in \eqref{eq:bulkcurrent} and the bulk quasi-particle currents in Eq.~\eqref{eq:bulkqpcurrent}. Outside the finite Hall sample, these currents simply vanish,
\begin{equation}\label{eq:Jn2}
\jb_n(x) = \langle\jbo_n(x)\rangle =
\begin{cases}
  \chi_n d\Af_n, & x\in \mathring\Lambda \\
  0, & x\not\in \Lambda\, .
\end{cases}
\end{equation}
An additional contribution to the current is localized on the boundary $\partial\Lambda$ of space-time. We can compute it as
\begin{equation}\label{eq:edcurrent}
  \jeo^\mu = \frac{\delta}{\delta a_\mu} (\Gamma_{\Lambda}[\Afb] + S_{\partial\Lambda}[{\bm \phi},\afb]) = \frac{1}{2}\sum_n Q_n (g^{\mu\nu}_n + \chi_n \epsilon^{\mu\nu}) (\partial_\nu\phi_n + \af_{n\nu})\, .
\end{equation}
Similar to the electric current in the bulk, the ``edge'' current \eqref{eq:edcurrent} can be decomposed into a sum of $N$ quasi-particle currents, $\jeo^\mu = \sum_n Q_n \jeo^\mu_n$, where
\begin{equation}\label{eq:edcurrents}
  \jeo_n^\mu = \frac{1}{2}(g^{\mu\nu}_n + \chi_n\epsilon^{\mu\nu})(\partial_\nu\phi_n + \af_{n\nu})\, .
\end{equation}

Inserting the chiral constraint \eqref{eq:motionedge1} or using the field equation for $\phi_n$ derived from the action \eqref{eq:effactionedge}, one finds from \eqref{eq:edcurrents} that
\begin{equation}\label{eq:divjn}
  \partial_\mu \jeo^\mu_n = \chi_n d\af_n\, . %= \chi_n Q_n da\, .
\end{equation}
The divergence in \eqref{eq:divjn} is canceled by the divergence of the corresponding bulk current Eq.~\eqref{eq:Jn2}. The total quasi-particle currents, containing both bulk, \eqref{eq:Jn2}, {\it and} edge, \eqref{eq:edcurrents}, contributions, are conserved.

\subsection{Quantization of the gapless edge modes}

In this section, we consider the quantization of the edge modes $\phi_n$ using the boundary action \eqref{eq:effactionedge}. For this purpose, it is convenient to shift the fields $\phi_n$ by appropriate functionals of the external gauge field $\af_{n}$. The field equation for $\phi_n$ derived from the action \eqref{eq:effactionedge} is
\begin{equation}\label{eq:motion3}
  g_n^{\mu\nu}\partial_\mu(\partial_\nu \phi_n + \af_{n\nu}) = \chi_n \epsilon^{\mu\nu}\partial_\mu\af_{n\nu}\, .
\end{equation}
The general solution to \eqref{eq:motion3} has the form
\begin{equation}\label{eq:defvarphi}
  \phi_n = p_n[\af_n] + \varphi_n\,,
\end{equation}
where $p_n[\af_n]$ is a gauge-dependent c-number field and the fluctuation field, $\varphi_n$, is the operator-valued solution of the wave equation,
\begin{equation}\label{eq:waveeq}
   g^{\mu\nu}_n\partial_\mu\partial_\nu\varphi_n = 0\,,
\end{equation}
on $\partial\Lambda = \partial\Omega\times\mathbb{R}$. The fields $\varphi_n$ are gauge-invariant and the corresponding action is simply given by
\begin{equation}\label{eq:freeboson}
  S_{\partial\Lambda}[{\bm \varphi}] = \frac{1}{4} \sum_n \int_{\partial\Lambda} d^2\xi\, g_n^{\mu\nu} \partial_\mu\varphi_n\partial_\nu\varphi_n \, .
\end{equation}
The free Bose fields $\varphi_n$ can now be canonically quantized in a standard fashion. Note that we do {\it not} fix boundary conditions for the fields $\varphi_n$. Canonical quantization yields the following commutation relations for the edge currents \eqref{eq:edcurrents} and \eqref{eq:edcurrent},
\begin{equation}
  [\jeo_n^0(\xi,t), \jeo_m^0(\eta,t)] = \delta_{n m}\chi_n i (2\pi)^{-1} \delta'(\xi-\eta)
\end{equation}
and
\begin{equation}
  [\jeo_n^0(\xi,t), \jeo^0(\eta,t)] = \chi_n Q_n i (2\pi)^{-1} \delta'(\xi-\eta)\, .
\end{equation}
Hence, the currents $\jeo_n$ are the generators of U(1) chiral Kac-Moody algebras \cite{GoddardOlive86}.

The field $p_n[\af_n]$ in \eqref{eq:defvarphi} is a particular solution to Eq.~\eqref{eq:motion3}. It includes the effect of the external gauge field $\af_n$ and is responsible for the correct gauge transformation of $\phi_n$, \eqref{eq:phitildetransf}. An explicit solution is given by
\begin{equation}\begin{split}
\label{eq:ourp}
  p_n[\af_n](\xi, t) = p_n^0&(\xi + \chi_n u_n (t-t_*))\\
   &+ \int^t_{t_*} d\bar t\, \{ \chi_n u_n \af_{n 1}(\xi + \chi_n u_n(t-\bar t), \bar t) - \af_{n 0}(\xi + \chi_n u_n(t-\bar t), \bar t) \}\,.
\end{split}\end{equation}
The function $p_n^0(\xi)$ in \eqref{eq:ourp} depends on the initial conditions at $t = t_*$ that we may specify for the particular solution, and it incorporates the effect of ``large'' gauge transformations not vanishing at the initial time $t_*$. A convenient choice is
\begin{equation}\label{eq:ourp0}
  p_n^0(\xi) = - \int^\xi_{\xi_*} d\bar\xi\, \af_{n 1}(\bar\xi,t_*)\,,
\end{equation}
where $\xi_*$ is some reference point on $\partial\Omega$; (More details are given in \ref{app:conscurr}). From \eqref{eq:ourp} and \eqref{eq:ourp0}, one can check that\footnote{Without loss of generality we consider gauge transformations with $\alpha(\xi_*,t_*) = 0$.}
\begin{equation}\label{eq:ptrans}
  p_n[\af_n + d\alpha_n](\xi,t) = p_n[\af_n](\xi,t) - \alpha_n(\xi,t)\,.
\end{equation}
Hence, $p_n$ transform correctly under gauge transformations [cf. Eqs.~\eqref{eq:defvarphi} and \eqref{eq:phitildetransf}].

In general, the function $p_n(\xi,t)$ is not single-valued on $\partial\Omega$ due to the first term in \eqref{eq:ourp}, $p_n^0(\xi)$. From \eqref{eq:ourp0}, the presence of magnetic fluxes or quasi-particle charges inside the boundary $\partial\Omega$ at time $t_*$ makes this term multi-valued. However, the second term in \eqref{eq:ourp} is single-valued on $\partial\Omega$ for arbitrary time-dependent gauge fields. In particular, this is true for an adiabatic increase of magnetic flux through the sample. This point will be of some importance in the discussion of the flux-dependence of tunneling currents between disjoint edge components in Sect.~\ref{sec:dynmod}.

\subsection{Conserved edge currents and quasi-particle vertex operators\label{sec:edgecurrents}}

In this section we construct quasi-particle operators localized on the boundary $\partial\Omega$. Before discussing these operators, we need to define the {\it conserved currents}, $\jeoh^\mu_n$, by
\begin{equation}\label{eq:conscurrents}
  \jeoh^\mu_n = \jeo^\mu_n - \chi_n \epsilon^{\mu\nu} \af_{n\nu}\, .
\end{equation}
For later purpose, we also define the conserved electric current on the edge as
\begin{equation}\label{eq:conscurrent}
  \jeoh^\mu = \jeo^\mu - \sum_n \chi_n Q_n  \epsilon^{\mu\nu} \af_{n\nu}\, .
\end{equation}
From Eq.~\eqref{eq:divjn}, it is clear that the currents $\jeoh^\mu_n$ are conserved, i.e., $\partial_\mu \jeoh^\mu_n = 0$. Note however that these currents are {\it gauge-dependent}.

Using \eqref{eq:ourp} and \eqref{eq:edcurrents}, the conserved currents, \eqref{eq:conscurrents}, can be written in terms of the shifted fields \eqref{eq:defvarphi} as\footnote{See also \ref{app:conscurr}.}
\begin{equation}\label{eq:conscurrents2}
  \jeoh^{\mu}_n = \frac{1}{2}(g^{\mu\nu}_n + \chi_n\epsilon^{\mu\nu})\partial_\nu\varphi_n + \chi_n \epsilon^{\mu\nu}\partial_\nu p_n\,.
\end{equation}
We see that the conserved currents only depend on one chiral component of the free Bose fields $\varphi_n$ and on the total derivative of the c-number field $p_n[\af_n]$. These facts will be useful shortly.

A quasi-particle operator is defined as a {\it vertex operator} in the following way:
\begin{equation}\label{eq:psiedge}\begin{split}
  \psi_{\bm{q}}(\xi, t) &:= \mathcal{N} exp\{ 2\pi i \sum_n q_n \int_{\gamma(\xi_*,\xi)} d\bar\xi^\mu\, \epsilon_{\mu\nu} \jeoh_n^\nu(\bar\xi) \}\,,
\end{split}\end{equation}
where $\gamma(\xi_*,\xi)$ is some path on the space-time boundary $\partial\Lambda$ joining the points $(\xi_*,t_*)$ and $(\xi,t)$, and $\bm{q} = (q_n)_{n=1}^N\in \mathbb{R}^N$ are real numbers. $\mathcal{N}$ is a normal-ordering prescription for the Bose fields $\varphi_n$; (In the following, we omit writing $\mathcal{N}$ in the vertex operator but understand it implicitly). The starting point $(\xi_*, t_*)$ of the path $\gamma$ is a reference point which acts as a {\it source of charges}. However, for $t_*\rightarrow -\infty$, its presence can be neglected in the expressions that follow.

\begin{figure}[!h]
\center
\includegraphics[width=0.5\textwidth]{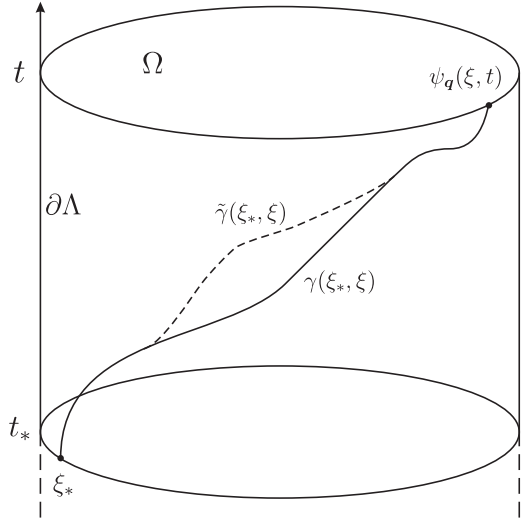}
\caption{The vertex operator $\psi_{\bm q}(\xi,t)$, \eqref{eq:psiedge}, is defined through a path $\gamma$ on $\partial\Lambda$ leading from a reference point $(\xi_*,t_*)$ to the point $(\xi,t)$. However, the vertex operator is invariant under continuous deformations of the path $\gamma(\xi_*,\xi)\mapsto\tilde\gamma(\xi_*,\xi)$, keeping the end-points fixed; it only depends on the homotopy class of the path.
\label{fig:cyllinder}
}
\end{figure}

The vertex operator \eqref{eq:psiedge} is invariant under continuous deformations of the contour $\gamma(\xi_*,\xi)$ with fixed endpoints on $\partial\Lambda$. This is due to the conservation of the currents, $\partial_\mu \jeoh_n^\mu = 0$. Hence, $\psi_{\bm{q}}(\xi,t)$ only depends on the homotopy class of the path $\gamma$ on $\partial\Lambda$; (see Fig.~\ref{fig:cyllinder}).

Under a gauge transformation [with $\alpha(\xi_*,t_*)=0$], the vertex operator transforms as
\begin{equation}\label{eq:psitrans}
  \psi_{\bm{q}}(\xi,t) \mapsto \psi_{\bm{q}}(\xi,t)\, exp\{ - 2\pi i \sum_n \chi_n q_n \alpha_n(\xi,t)\}\, .
\end{equation}
This gauge variation suggests that $\psi_{\bm{q}}(\xi,t)$ deposits the charges $(\chi_n q_n)$, and an electric charge
\begin{equation}
  Q_{el}(\bm{q}) = \sum_n \chi_n Q_n q_n
\end{equation}
at the point $\xi$ of the edge.

The edge charges deposited by $\psi_{\bm{q}}(\xi,t)$ can be obtained from the operator viewpoint. Using \eqref{eq:conscurrents2} and \eqref{eq:ourp}, the operators of conserved charge are given by
\begin{equation}\label{eq:quasicharge}
  \hat q_n = \int_{\partial\Omega} d\xi\, \jeoh_n^{0}(\xi)  = \chi_n \int_0^L d\xi\, \{ \partial_\xi p_n^0(\xi) + \frac{1}{2} ( \varphi'_n + \chi_n u_n^{-1} \dot\varphi_n) \}\,.
\end{equation}
The operator of conserved electric charge is
\begin{equation}\label{eq:conservedelectricchrg}
  \hat Q = \sum_n Q_n \hat q_n\, .
\end{equation}
The commutators of charge- and vertex operators are given by
\begin{equation}\label{eq:charge}
  [\hat q_n, \psi_{\bm{q}}(\xi,t)] = \chi_n q_n\, \psi_{\bm{q}}(\xi,t)
\end{equation}
and
\begin{equation}\label{eq:commut}
  [\hat Q, \psi_{\bm{q}}(\xi,t)] = Q_{el}({\bm{q}})\, \psi_{\bm{q}}(\xi,t)\,,
\end{equation}
which agrees with the previous conclusions.

Besides the electric charge, the {\it scaling dimension}, $g$, is an important property of the vertex operator. It is defined as
\begin{equation}
  g({\bm q}) = \sum_n q_n^2\, .
\end{equation}
We will see that the scaling dimension $g$ determines the exponent of the two-point function of vertex operators.

\subsubsection{Gauge-invariant (bare) vertex operators}

Using the expression \eqref{eq:conscurrents2} for the conserved currents and light-cone coordinates,
\begin{equation}\label{eq:lightcone}
  x_n(\xi,t) = \chi_n u_n^{-1}\xi + t\,,
\end{equation}
the vertex operator, Eq.~\eqref{eq:psiedge}, can be written as
\begin{equation}\label{eq:psiedge2}\begin{split}
  \psi_{\bm{q}}(\xi,t) = exp\{ &2\pi i \sum_n q_n \chi_n \int_{\gamma(\xi_*,\xi)} d\bar\xi^\mu\, \partial_\mu p_n(\bar\xi) \}\\
   &exp \{ 2 \pi i\, \sum_n q_n \chi_n \int^{x_n(\xi)} dx_n\, \partial^{n}\varphi_n \}\,.
\end{split}\end{equation}
$\partial^n = \frac{1}{2}(\partial_0 + \chi_n u_n \partial_1)$ is the derivative with respect to the coordinate \eqref{eq:lightcone}. The second factor in this product,
\begin{equation}\label{eq:vertexop}
  V_{\bm{q}}(\xi,t) := exp \{ 2 \pi i\, \sum_n q_n \chi_n \int^{x_n(\xi,t)} dx_{n}\, \partial^{n}\varphi_n \}\,,
\end{equation}
is gauge-invariant. We call $V_{\bm q}(\xi,t)$ the {\it bare} vertex operator. If all the modes $\varphi_n$ had the same propagation speeds $u_n$ and -directions $\chi_n$, \eqref{eq:vertexop} would be a {\it chiral} vertex operator \cite{Difrancesco97}.

\subsection{Superselection sectors of edge states}\label{sec:chargesectors}

In the previous section, we have introduced vertex operators $\psi_{\bm q}(\xi,t)$ that create states of charge $(\chi_n q_n)$, as expressed by Eq.~\eqref{eq:charge}. The (half-)monodromies of the localized bulk states, see \eqref{eq:monodromy} and \eqref{eq:chargestatistics}, correspond to commutators of vertex operators. It is straightforward to find that
\begin{equation}
  \psi_{{\bm q}_1}(\xi,t)  \psi_{{\bm q}_2}(\eta,t) = \psi_{{\bm q}_2}(\eta,t) \psi_{{\bm q}_1}(\xi,t)\, \exp \{ i \pi\, \text{sgn}(\xi-\eta)\, {{\bm q}_1} \cdot {{\bm q}_2} \}\,,
\end{equation}
for $\xi,\eta>0$. Our classification of physical states of the bulk of an incompressible Hall fluid also applies to vertex operators, and the vectors ${\bm q}$ in the definition \eqref{eq:psiedge} must correspond to the sites of the charge lattice, ${\bm q}\in \Gamma_e^*\oplus\Gamma_h^*$.

The Hilbert space of edge states decomposes into superselection sectors for each connected component of $\partial\Omega$. These sectors are labeled by the eigenvalues of the operator $\hat q_n$, given in Eq.~\eqref{eq:quasicharge}. Let ${\bm q} = (q_n)$ be a vector of charges. The ground state in a charge sector $[{\bm q}]$ is denoted by
\begin{equation}\label{eq:edgestate}
  |{\bm q}\rangle \in [{\bm q}]\,,
\end{equation}
with
\begin{equation}\label{eq:chargestate}
  \hat q_n |{\bm q}\rangle = \chi_n q_n |{\bm q}\rangle\, .
\end{equation}
From \eqref{eq:charge}, we have that
\begin{equation}
  \psi_{\bm q'}|{\bm q} \rangle\in [{\bm q} + {\bm q}']\, .
\end{equation}
Therefore, similar to the vertex operators $\psi_{\bm q}$, the sectors $[{\bm q}]$ are labeled by sites on the dual lattice, ${\bm q}\in \Gamma_e^*\oplus\Gamma_h^*$. Superselection sectors of edge states are in one-to-one correspondence to sectors of bulk states that were discussed in Section \ref{sec:bulkstates}.

\subsubsection{Charge sectors and a zero mode}\label{sec:chargezero}

From \eqref{eq:quasicharge} and \eqref{eq:chargestate}, we have that
\begin{equation}\label{eq:chargeop}
  \chi_n \hat q_n |{\bm q}\rangle = \int_0^L d\xi\, \{ \partial_\xi p^0_n + \frac{1}{2} ( \varphi'_n + \chi_n u_n^{-1} \dot\varphi_n) \} |{\bm q}\rangle = q_n |{\bm q}\rangle\, .
\end{equation}
Let us first consider the c-number contribution to the edge charge $q_n$. From \eqref{eq:ourp0} we find that
\begin{equation}\label{eq:chargeopc}
  \int_0^L d\xi\, \partial_\xi p^0_n(\xi) = -\int_0^L d\xi\, \af_{n 1}(\xi,t_*) =: \Phi_{* n}\, .
\end{equation}
Up to a sign, $\Phi_{* n}$ is the flux through, or quasi-particle charge in channel $n$ inside the edge component $\partial^{(k)}\!\Omega$ under consideration, at the initial time~$t_*$. Let us assume that the fluctuation of the electromagnetic gauge potential, $a_1$, vanishes at the initial time $t = t_*$. Then, on the one hand, if $\partial^{(k)}\!\Omega$ encloses a {\it hole} in the quantum Hall fluid, $\Phi_{* n} = 0$, because only the bulk supports localized quasi-particles. On the other hand, if we consider the outer edge component of the sample then $\Phi_{* n}$ is the total localized bulk charge; (see Fig.~\ref{fig:sample}).

The second contribution to the edge charge $q_n$ in Eq.~\eqref{eq:chargeop} comes from the eigenvalue of the operator
\begin{equation}\label{eq:zeromode}
  \hat \pi_n^0 = \frac{1}{2} \int_0^L d\xi\, (\varphi'_n + \chi_n u_n^{-1} \dot\varphi_n)\, .
\end{equation}
In conformal field theory \cite{Difrancesco97}, $\hat\pi_n^0$ is conventionally called a {\it zero mode} of the massless field $\varphi_n$. The zero mode $\hat \pi_n^0$ and the c-number $\Phi_{* n}$ are two independent quantities; the former is a genuine property of the edge state, while the latter characterizes the inside of $\partial^{(k)}\!\Omega$. Therefore, one may think that it would be more convenient to use the zero-mode eigenvalue $\pi_n^0$ to label the edge states. However, our choice of Eq.~\eqref{eq:chargeop} makes apparent the duality between the eigenvalues of \eqref{eq:zeromode} and the bulk charges \eqref{eq:chargeopc}. Instead of considering non-zero eigenvalues $\pi_n^0$, we can equally well thread additional {\it virtual $\Af_n$-fluxes} through the inside of the boundary component. That is, if $\partial^{(k)}\!\Omega$ encloses a {\it hole} in the sample we thread the flux $\tilde \Phi_{* n} = \pi_n^0 = q_n$, instead of $\Phi_{* n} = 0$, through the hole. If $\partial^{(k)}\!\Omega$ is the outer edge of the sample we thread the flux $\tilde \Phi_{* n} = -\Phi_{* n} - \pi_n^0 = - q_n$, instead of $-\Phi_{* n}$, through the bulk. %The charge $q_n$ is the same in both cases.

\subsubsection{Magnetic flux through the Hall sample and charge sectors}

Equations \eqref{eq:chargeop} and \eqref{eq:chargeopc} tell us that the charge $q_n$ of an edge state depends on the bulk flux $\Phi^n_*$ {\it at an initial time $t_*$}. In particular, this means that the charge eigenvalue $q_n$ remains constant even for time-dependent magnetic fluxes through the sample. Furthermore, the charge $q_n$ is invariant when localized bulk charges are moved in space by some external agent.\footnote{When a bulk charge is brought close to the edge by the external agent, it is reasonable to expect a jump in the bulk flux $\Phi^n_*$ and a corresponding change of the edge state such that $q_n$ is constant in \eqref{eq:chargeop}, i.e., a corresponding change in the zero-mode eigenvalue $\pi_n^0$. However, this process is not in the scope of our approach.} This is in contrast to the {\it physical} edge charges, which change when the magnetic flux through the sample is changed. The physical edge charges are discussed next.

\subsubsection{Electric charge localized on the edge}

The physical electric current density on some edge component $\partial^{(k)}\!\Omega$ is given by the operator $\jeo^\mu$, and {\it not} by the conserved current, $\jeoh^\mu$. By \eqref{eq:conscurrent}, the operator of conserved charge $\hat Q^{(k)}$ is related to the electric charge operator, $Q^{(k)}_{el}$, through
\begin{equation}\label{eq:fluxcharge}
  Q^{(k)}_{el} = \hat Q^{(k)} + \Phi_{\partial^{(k)}\!\Omega}(t)\,,
\end{equation}
where $\Phi_{\partial^{(k)}\!\Omega}(t)$ is the total flux through the surface enclosed by $\partial^{(k)}\!\Omega$ {\it at the time $t$}. It is given by
\begin{equation}\label{eq:chargechange}\begin{split}
  \Phi_{\partial^{(k)}\!\Omega}(t) &= \sigma \Phi^{\text{em}}_{\partial^{(k)}\!\Omega}(t) + \sum_n \chi_n Q_n \Phi_n(t)\\
   &= \sigma \int_{\partial^{(k)}\!\Omega} d\xi\, a_1(\xi, t)  + \sum_n \chi_n Q_n \int_{\Omega} d^2x\, \bar{\jb}_n^0({\bm x},t)\, .
\end{split}\end{equation}
Of course, if $\partial^{(k)}\!\Omega$ encircles a hole in the sample, then the last term in \eqref{eq:chargechange} is absent. Let us repeat that the electric charge on some edge component is, in general, {\it time dependent}. This is due to the non-conservation of the currents \eqref{eq:divjn}.

Let us summarize these facts: While the conserved edge charge operator $\hat Q$ has a discrete spectrum and the corresponding eigenstates are insensitive to time-dependent gauge fields, the electric charge $\langle Q_{el}\rangle$ can take arbitrary real values. The electric charge on a boundary component may change continuously when the external gauge fields are time-dependent.

\subsection{Two-point function of vertex operators}

To compute the tunneling current between boundary components in the next section, we will make use of the two-point function of quasi-particle operators localized on the edge. From \eqref{eq:psiedge2}, we have that
\begin{equation}\label{eq:psicorrelator}\begin{split}
  \langle {\bm q}'| &\psi_{\bm{q}}(\xi,t) \psi_{-\bm{q}}(\xi',t')|{\bm q}'\rangle\\
   &= exp\{ 2\pi i \sum_n q_n \chi_n \int_{\gamma(\xi,\xi')} d\bar\xi^\mu\, \partial_\mu p_n[\af_n](\bar\xi) \}\, \langle   V_{\bm{q}}(\xi,t) V_{-\bm{q}}(\xi',t') \rangle\,.
\end{split}\end{equation}
In \eqref{eq:psicorrelator}, the path $\gamma(\xi,\xi')$ leads from point $(\xi,t)$ to point $(\xi',t')$ on $\partial\Lambda$. Note that the reference point $\xi_*$ in the definition of the vertex operator drops out of the two-point function. However, the homotopy class of the path $\gamma$ enters the {\it phase} of the correlator. In Sect.~\ref{sec:ohmiccontact} we will see that the presence of an {\it Ohmic contact} on the edge $\partial\Omega$ essentially singles out one class of paths for which the two-point function is non-zero.

For $(\xi,t)\in\partial\Lambda = \mathbb{R}^2$ (on the infinite plane), the correlator of bare vertex operators is readily computed from \eqref{eq:freeboson} as
\begin{equation}\label{eq:barecorrelator}\begin{split}
  \langle V_{\bm q}(\xi,t) V_{-\bm q}(\xi',t')\rangle_{\mathbb{R}^2} &= \prod_n \left[ \frac{ i \epsilon}{u_n x_n(\xi-\xi',t-t')}\right]^{q_n^2}\\
   &= \prod_n \left[ \frac{i \epsilon}{(t-t')u_n + (\xi-\xi')\chi_n }\right]^{q_n^2}\,,
\end{split}\end{equation}
where $\epsilon$ is a cutoff length scale. The correlator at finite temperature $T$ can be obtained from \eqref{eq:barecorrelator} by a conformal transformation \cite{Shankar90,Difrancesco97}, applied to each term in the product over $n$ separately. Under the conformal transformation
\begin{equation}
  x_n \mapsto (2\pi T)^{-1} \log x_n
\end{equation}
in the light-cone coordinates $x_n$, \eqref{eq:lightcone}, one finds that
\begin{equation}\label{eq:barecorrelatorT}\begin{split}
  \langle V_{\bm q}(\xi,t) V_{-\bm q}(\xi',t')\rangle_{\mathbb{R}^2} \mapsto &\langle V_{\bm q}(\xi,t) V_{-\bm q}(\xi',t')\rangle_{T}\\
   &= \prod_n \left[ \frac{ \pi T\, i \epsilon }{u_n \sinh \{ \pi T x_n(\xi-\xi',t-t') \}} \right]^{q_n^2}\, .
\end{split}\end{equation}

We emphasize that the state of the edge, $|q_n'\rangle$, enters the two-point function through the {\it first term} in the product in \eqref{eq:psicorrelator}: The edge charges $q_n'$ appear as additional sources in the fields $\Af_n$ in the inside of the boundary component $\partial^{(k)}\!\Omega$ under consideration. This follows from our discussion after Eq.~\eqref{eq:chargeop}. Note that the (local) phase of the two-point function is gauge-dependent and, therefore, unobservable. However, we will see later that the edge charge, as a {\it global} property of the boundary component, may nevertheless enter physically observable quantities.

For later purpose, let us compute the phase of the two-point function \eqref{eq:psicorrelator} for a special choice of gauge fields. When $\af_{n 1}(\xi,t) = \af_{n 1}^c(\xi,t_*)$ is time-independent and $\af_{n 0}(\xi,t) = \af_{n 0}^c(0,t)$ is independent of space, then, from \eqref{eq:ourp} and \eqref{eq:ourp0}, one finds that
\begin{equation}\label{eq:pnspecial}
  \partial_\mu p_n[\af_n^c](\xi,t) = - \af_{n \mu}^c(\xi,t)\, .
\end{equation}
Using Eq.~\eqref{eq:pnspecial}, the phase of the two-point function \eqref{eq:psicorrelator} is then given by
\begin{equation}\label{eq:pnspecialint}
  \int_{\gamma(\xi,\xi')} d\bar\xi^\mu\, \partial_\mu p_n[\af_n^c](\bar\xi,\bar t) = - \int_t^{t'} d\bar t\, \af_{n 0}^c(0, \bar t) - \int_\xi^{\xi'} d\bar\xi\, \af_{n 1}^c(\bar \xi,t_*)\, .
\end{equation}

\section{Inter-edge charge transport}\label{sec:interedgetransport}

\begin{figure}[!h]
\center
\includegraphics[width=0.6\textwidth]{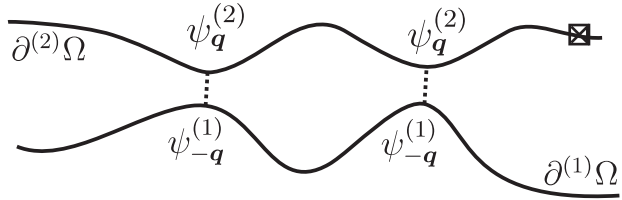}
\caption{Two tunneling constrictions between two boundary components of the quantum Hall edge. Each edge component is maintained at a chemical potential by an Ohmic contact (crossed box on upper edge).
\label{fig:tunneling}
}
\end{figure}

In this section, we consider tunneling processes between two connected boundary components of a quantum Hall sample. We want to investigate transport properties of tunneling junctions. More precisely, our aim is to calculate the electric current through junctions as a function of the state of the quantum Hall sample, of changes in magnetic flux through the sample, and of the bias voltage across tunneling constrictions. Physically, tunneling junctions can be created by deforming the quantum Hall edge to form sufficiently narrow {\it constrictions}; see, e.g., \cite{FrohlichPedrini2001}, (Fig.~\ref{fig:tunneling}).

The analysis of inter-edge charge transport calls for two additional concepts in the description of the quantum Hall sample. First, we need to introduce the notion of an {\it Ohmic contact}, which maintains the edge at some given voltage. Second, we have to define {\it tunneling operators} modeling the presence of constrictions.

\subsection{Ohmic contact}\label{sec:ohmiccontact}

We define the Ohmic contact as a location on the edge component $\partial^{(k)}\!\Omega$ where electrons are removed or added in order to maintain a fixed chemical potential. Let $L = |\partial^{(k)}\!\Omega|$ be the total length of the edge component. The presence of an Ohmic contact strongly affects the physics close to the contact. However, far from the contact, the correlators are expected to be essentially unaffected by it. This means that the two-point functions of bare vertex operators in the presence of an Ohmic contact continue to be (essentially) given by Eqs.~\eqref{eq:barecorrelator} and \eqref{eq:barecorrelatorT}, when the distance of the arguments, $\xi$ and $\xi'$, to the Ohmic contact is $O(L)$. In particular, this means that
\begin{equation}\label{eq:condition}
  |\xi-\xi'|\ll L
\end{equation}
for our analysis to be valid.

Furthermore, we shall assume that two-point functions of vertex operators, \eqref{eq:psicorrelator}, that have a path $\gamma$ {\it crossing the Ohmic contact at some time}, are small and can be neglected. As a result, in a given gauge, an Ohmic contact fixes the phases of all two-point functions of vertex operators.

\subsection{Tunneling operator}\label{sec:tunnelingoperator}

In this section, we define gauge-invariant {\it tunneling operators} that model the presence of narrow constrictions between two boundary components of a quantum Hall sample. The tunneling operator is supposed to transfer quasi-particles from one boundary component to another one.

The two boundary components involved in a tunneling process enclose the same quantum Hall fluid. As we discussed in Sect.~\ref{sec:GI}, this imposes constraints on the field theories describing these boundary components.\footnote{We are considering here the case of a sharp, ``unreconstructed'' quantum Hall edge. More complicated boundary structures may change the transport properties of the edge in a nontrivial manner.} In particular, the number of modes $N$, the electric charge parameters $Q_n$, and the chiralities $\chi_n$ must coincide. However, the propagation speeds $u_n$ generally differ on the two boundary components. Let us denote the two edge components by $\partial^{(k)}\!\Omega$, $k = 1, 2$. The action describing the dynamics of the relevant edge degrees of freedom, $\phi_n^{(k)}$, is $S_{\partial^{(1)}\!\Omega} + S_{\partial^{(2)}\!\Omega}$, where each $S_{\partial^{(k)}\!\Omega}$ is given by Eq.~\eqref{eq:effactionedge}. The corresponding Hamiltonians are denoted by $H^{(k)}$; (we do not need their explicit expressions).

We model the constrictions by adding a tunneling operator, $T$, as a perturbation to the total Hamiltonian,
\begin{equation}\label{eq:totalham}
  H = H_0 + T = H^{(1)} + H^{(2)} + T\,.
\end{equation}
The tunneling operator couples the two boundary components through terms $T \propto \psi_{{\bm{q}}^{(1)}}^{(1)} \psi_{{\bm{q}}^{(2)}}^{(2)}$. Total charge conservation requires that ${\bm{q}}^{(1)} + {\bm{q}}^{(2)} = 0$.

In the scaling limit under consideration, only the most relevant perturbations, in the renormalization-group sense, need to be retained in the tunneling operator. A perturbation $\propto \psi_{\bm q}$ is more relevant the {\it smaller} the corresponding scaling dimension, $g = \sum_n q_n^2$; see, e.g., \cite{WenIJMP92,Yoshii89}. Therefore, in the scaling limit, quasi-particle tunneling operators with ${\bm q}\in \Gamma^*_e\otimes\Gamma^*_h$ and $\sum_n q_n^2 = \sum_n q_{e n}^2 + \sum_n q_{h n}^2$ as small as possible are favored.

We suppose that quasi-particles of charge $\bm{q}$ may tunnel at {\it different locations}, $\xi_p$, of the edge components. The gauge-invariant tunneling operator is then given by
\begin{equation}\label{eq:Top}
  T = \sum_p t_p\, \psi^{(1)}_{\bm{q}}(\xi_p^{1}) \psi^{(2)}_{-\bm{q}}(\xi_p^{2}) \exp\{-2\pi i \sum_n \chi_n q_n \int_{\xi_p^{1}}^{\xi_p^{2}} \Af_n \} + h.c.
\end{equation}
The constants $t_p$ are the tunneling amplitudes. In the following, we assume that the constrictions are quite narrow, $\xi_p^{1}\simeq \xi_p^{2}$, such that the phase factor in the tunneling operator \eqref{eq:Top} can be neglected.\footnote{The phase factor in the tunneling operator \eqref{eq:Top} may be taken into account without difficulty. We simply omit it here to make our notation more compact. This phase will be important in Sect.~\ref{sec:dragging} when discussing the effect of bulk quasi-particles that are dragged around by an external agent.}

Note that, in principle, the tunneling operator \eqref{eq:Top} depends on the homotopy classes of paths $\gamma$ used in the definition of the vertex operators $\psi^{(k)}_{\bm q}$. However, the Ohmic contacts effectively cut open the boundary space-time cylinders $\partial^{(k)}\!\Lambda$ along their locations. As discussed in the previous section, two-point functions of vertex operators are assumed to vanish for paths $\gamma$ crossing the Ohmic contact. Thus, the vertex operators used in \eqref{eq:Top} are uniquely specified by the locations of the Ohmic contacts.

\subsection{Perturbative tunneling current}

Let us compute the electric tunneling current flowing from one boundary component to the other through the constrictions. In the absence of tunneling, the charges $\hat Q^{(k)}$, Eq.~\eqref{eq:conservedelectricchrg}, are separately conserved on each edge. At the operator level, this means that
\begin{equation}\label{eq:chargecons}
  %\frac{d}{dt} \hat Q^{(k)} =
  \frac{\partial}{\partial t}\, \hat Q^{(k)} + \frac{i}{2\pi}\, [H^{(k)},\hat Q^{(k)}] = 0\, .
\end{equation}
The tunneling current is the time derivative of the electric charge on one edge, say $(k)=(1)$. Using Eqs.~\eqref{eq:commut}, \eqref{eq:fluxcharge}, and \eqref{eq:totalham}-\eqref{eq:chargecons}, it is found to be given by
\begin{equation}\label{eq:tunnelingcurrent}\begin{split}
  I &= \frac{d}{dt} Q^{(1)}_{el} = \frac{d}{dt} \hat Q^{(1)} + \frac{d}{dt} \Phi_{\partial^{(1)}\!\Omega} = \frac{i}{2\pi}\, [\hat Q^{(1)}, T] + \frac{d}{dt} \Phi_{\partial^{(1)}\!\Omega}\\
   &= \frac{i}{2\pi}\, Q_{el}({\bm{q}}) \sum_p \{  t_p\, \psi^{(1)}_{\bm{q}}(\xi_p^{1}) \psi^{(2)}_{-\bm{q}}(\xi_p^{2}) - t_p^*\, \psi^{(1)}_{-\bm{q}}(\xi_p^{1}) \psi^{(2)}_{\bm{q}}(\xi_p^{2})\} + \frac{d}{dt} \Phi_{\partial^{(1)}\!\Omega}\,.
\end{split}\end{equation}

The term $\frac{d}{dt} \Phi_{\partial^{(1)}\!\Omega}$ in \eqref{eq:tunnelingcurrent} corresponds to charging effects of the edge resulting from the variation of magnetic flux through the surface enclosed by $\partial^{(1)}\!\Omega$. According to the non-conservation of edge currents, \eqref{eq:divjn}, charge flows through the bulk to the edge when the external fields are varied. However, we are interested in the current flowing from one edge to the other via the constrictions, so we omit this term in the following.

Next, we apply time-dependent perturbation theory in the tunneling amplitudes $t_p$ to computing the tunneling currents. The free time evolution is generated by the Hamiltonian $H_0 = H^{(1)} + H^{(2)}$. The Dyson series for the current operator is
\begin{equation}\label{eq:currentpert}
  I(t) = e^{i t H} I e^{-i t H} = I_0(t) + i (2\pi)^{-1} \int_0^t dt'\, [T_{0}(t'),I_0(t)] + \ldots
\end{equation}
where
\begin{equation}
  A_0(t) =  e^{i t H_0} A e^{-i t H_0 }
\end{equation}
is the free time evolution.

We now take the expectation value of the current in a given initial equilibrium state, $\langle(\cdot)\rangle$, of the quantum Hall fluid. Using expression \eqref{eq:tunnelingcurrent} for the current operator, $\langle I_0(t)\rangle = 0$, and $\langle\psi_{\bm q}\psi_{\bm q}\rangle = 0$, one finds that
\begin{equation}\begin{split}\label{eq:current0}
  \langle I(t)\rangle = \frac{{\tilde e}}{(2\pi)^{2}} \sum_{p q} \{ &t_p t_q^* \int_{0}^t dt'\, \langle [\psi^{(1)}_{\bm{q}}(\xi_p^{1},t') \psi^{(2)}_{-\bm{q}}(\xi_p^{2},t'), \psi^{(1)}_{-\bm{q}}(\xi_q^{1},t) \psi^{(2)}_{\bm{q}}(\xi_q^{2},t) ]\rangle\\ &+ c.c. \}+ O(t_{\cdot}^4)\,,
\end{split}\end{equation}
where ${\tilde e} = Q_{el}({\bm q})$ is the electric charge of the quasi-particle that tunnels at the constrictions. Next, we discuss the choice of an appropriate initial state $\langle(\cdot)\rangle$.

\subsubsection{Initial state for perturbation theory}\label{sec:initialstate}

In a first approach, we suppose that $\langle(\cdot)\rangle$ is a thermal equilibrium state {\it in a given charge superselection sector of the edges}, as introduced in Sect.~\ref{sec:chargesectors}. That is, the sector is specified by the charges on all boundary components $\partial^{(k)}\!\Omega$ and by the localized bulk charges at the points $z_m\in \Omega$. We denote such a sector by
\begin{equation}\label{eq:chargesector}
  [ \{{\bm q}^{(k)}\}; \{({\bm q},z)^m\} ]\,.
\end{equation}

However, inter-edge tunneling processes modify the charges on the edge components involved in the tunneling process. We have that
\begin{equation}\begin{split}
  \psi^{(1)}_{\bm q}\psi^{(2)}_{-\bm q}\, &[ {\bm q}^{(1)}, {\bm q}^{(2)}, \{{\bm q}^{(k)}\}_{k\notin \{1, 2\}}; \{({\bm q},z)^m\} ]\\
  &= [ {\bm q}^{(1)}+{\bm q}, {\bm q}^{(2)}-{\bm q}, \{{\bm q}^{(k)}\}_{k\notin\{1, 2\}}; \{({\bm q},z)^m\} ]\, .
\end{split}\end{equation}
An appropriate initial state belongs therefore to a subspace,
\begin{equation}\label{eq:delocalizedsector}\begin{split}
  &[ {\bm q}^{(1)}, {\bm q}^{(2)}, \{{\bm q}^{(k)}\}_{k\notin\{1, 2\}}; \{({\bm q},z)^m\} ]_{\bm q}\\
  &= \bigoplus_{M} [ {\bm q}^{(1)} + M {\bm q}, {\bm q}^{(2)} - M {\bm q}, \{{\bm q}^{(k)}\}_{k\notin \{1, 2\}}; \{({\bm q},z)^m\} ]\, .
\end{split}\end{equation}
The sector \eqref{eq:delocalizedsector} is an invariant subspace for the tunneling operator $T$, \eqref{eq:Top}, i.e.,
\begin{equation}\begin{split}
  T &[ {\bm q}^{(1)}, {\bm q}^{(2)}, \{{\bm q}^{(k)}\}_{k\notin\{1, 2\}}; \{({\bm q},z)^m\} ]_{\bm q}\\
   &= [ {\bm q}^{(1)}, {\bm q}^{(2)}, \{{\bm q}^{(k)}\}_{k\notin\{1, 2\}}; \{({\bm q},z)^m\} ]_{\bm q}\, .
\end{split}\end{equation}

In the following, we choose an initial state, $\langle(\cdot)\rangle$, in a charge superselection sector \eqref{eq:chargesector}. We find that the tunneling current through a single constriction is {\it independent} of the charge. The {\it interference term} in the current for tunneling through two or more constrictions does, however, depend on the charge sector. The current resulting from an initial state in the subspace \eqref{eq:delocalizedsector} is then obtained by taking an appropriate average of the interference term over different charge sectors.

\subsubsection{Interference phase}

Using the representation \eqref{eq:psiedge2} for the vertex operators and an initial state $\langle(\cdot)\rangle$ in a charge sector \eqref{eq:chargesector}, the expression \eqref{eq:current0} for the tunneling current can be further simplified. It is given by
\begin{equation}\begin{split}\label{eq:current2}
  \langle I(t)\rangle& = \frac{{\tilde e}}{(2\pi)^{2}} \sum_{p q} \{ t_p t_q^*\\ &\int_{0}^t dt' e^{ 2\pi i\, \Phi^{pq}(t',t) }\,
  \langle [ V^{(1)}_{\bm{q}}(\xi_p^{1}, t') V^{(2)}_{-\bm{q}}(\xi_p^{2},t') , V^{(1)}_{-\bm{q}}(\xi_q^{1},t) V^{(2)}_{\bm{q}}(\xi_q^{2},t)]\rangle + c.c.\}\, .
\end{split}\end{equation}
Here, we have introduced the {\it interference phase} $\Phi^{pq}(t',t)$ given by
\begin{equation}\begin{split}
\label{eq:interferencephase}
\Phi^{pq}(t',t) = \sum_n \chi_n q_n \{ \oint_{\gamma_{pq}(t',t)} d\bar\xi^\mu\, \partial_\mu p_n[\af_n](\bar\xi) - \int_{\xi_q^1}^{\xi_q^2} \Af_n(t') - \int_{\xi_p^2}^{\xi_p^1} \Af_n(t)\}\,.
\end{split}\end{equation}
The line integrals of the bulk fields $\Af_n$ appear in \eqref{eq:interferencephase} because we choose to reintroduce the phase factors in the tunneling operator \eqref{eq:Top}. The loop $\gamma_{pq}(t',t)$ is the {\it interference contour}. The explicit expressions of the functions $p_n[\af_n]$ for arbitrary gauge fields $\af_n$ are as given in Eqs.~\eqref{eq:ourp} and \eqref{eq:ourp0}.

The interference contour $\gamma_{pq}(t',t)$ starts at the point $\xi_p^{1}$ at time $t'$ and ends at $\xi_q^{1}$ at time $t$, along the first edge; then it crosses to the point $\xi_q^{2}$ on the second edge and from there to $\xi_p^{2}$ at time $t'$. Finally, it returns to the initial point on the first edge. Which one of the two alternative paths connecting $\xi_p^{k}$ and $\xi_q^{k}$ on each edge component, $(k)=(1),(2)$, must be taken is determined by the location of the Ohmic contact: The path must be chosen so as {\it not} to cross the Ohmic contact at any time. Other paths do not make any contribution to the tunneling current. This follows from the definition of the tunneling operator and from our discussion in Sects.~\ref{sec:ohmiccontact} and \ref{sec:tunnelingoperator}. (See also Figs.~\ref{fig:MZ1} and \ref{fig:FP1}.)

Note that the interference phase $\Phi^{pq}(t',t)$, Eq.~\eqref{eq:interferencephase}, depends on the {\it state} of the quantum Hall fluid. In particular, it depends on the {\it charge sectors} of the two boundary components in the initial state~$\langle(\cdot)\rangle$. The two-point functions of {\it bare} vertex operators in \eqref{eq:current2}, however, only depend on the temperature, but not on the charge sector of the initial state.

\subsubsection{Interference phase for static gauge fields and inter-edge bias voltage}

We are interested in the stationary value of the interference current \eqref{eq:current2} that is reached after some equilibration time $t_{eql}$. We suppose that the equilibration time is sufficiently short, such that the gauge fields on the boundary, $\af_n$, are essentially time-independent, for $0 \leq t \leq t_{eql}$. Furthermore, we introduce a dc voltage difference $U$ between the two boundary components by letting
\begin{equation}\label{eq:bias}
  a_0^{(1)} - a_0^{(2)} = U
\end{equation}
and $a_{n0}=0$ for all times. In this special case, the expressions \eqref{eq:pnspecial} and \eqref{eq:pnspecialint} for $p_n[\af_n]$ can be used to evaluate the interference phase \eqref{eq:interferencephase}. We find that
\begin{equation}\label{eq:interferencephase2}
  \Phi^{pq}(t',t) = (t-t')\, {\tilde e} U  - \sum_n \chi_n q_n \oint_{\gamma_{pq}} d\xi\, \af_{n 1}(\xi, 0) \,,
\end{equation}
where $\gamma_{pq}$ is the interference path shifted to the initial time $t=0$; (see Figs.~\ref{fig:MZ1} and \ref{fig:FP1}). The last, time-independent term in \eqref{eq:interferencephase2} is simply the total quasi-particle charge located inside the interference contour at the initial time $t=0$. We define it as
\begin{equation}\label{eq:interferencephase3}
  \Phi^{pq}_* =  - \sum_n \chi_n q_n \oint_{\gamma_{pq}} d\xi\, \af_{n 1}(\xi, 0) \,.
\end{equation}
From \eqref{eq:interferencephase3}, it is clear that $\Phi^{pq}_* = -\Phi^{qp}_*$ and $\Phi^{qq}_*=0$. From our discussion in Sect.~\ref{sec:chargezero} it follows that, not only do the localized bulk charges contribute to \eqref{eq:interferencephase3}, but also the edge charges. This point will be discussed in more detail later on.

We now turn back to an evaluation of the expression \eqref{eq:current2} for the tunneling current. We set $s=t'-t$ and take the limit $t\rightarrow\infty$ in \eqref{eq:current2}. Using time-translational invariance of the two-point function of bare vertex operators, Eq.~\eqref{eq:barecorrelator}, we finally obtain the following expression for the tunneling current:
\begin{equation}\begin{split}\label{eq:retcf}
  \langle I\rangle =
  %&\frac{Q_{el}}{(2\pi)^{2}} \sum_p t_p^2\int_{0}^\infty dt\{ e^{ 2\pi i\, Q_{el} V t }\,
  %\langle [ V^{(1)}_{\bm{q}}(\xi_p^{1},t) V^{(2)}_{-\bm{q}}(\xi_p^{2},t) , V^{(1)}_{-\bm{q}}(\xi_p^{1}) V^{(2)}_{\bm{q}}(\xi_p^{2})]\rangle + c.c.\}\\
  &\frac{{\tilde e}}{(2\pi)^{2}} \sum_{p q} \{ t_p t_q^*\, e^{2\pi i \Phi^{pq}_*}\\
  &\int_{-\infty}^0 ds\, e^{ -2\pi i\, {\tilde e} U s }\, \langle [ V^{(1)}_{\bm{q}}(\xi_p^{1},s) V^{(2)}_{-\bm{q}}(\xi_p^{2},s) , V^{(1)}_{-\bm{q}}(\xi_q^{1},0) V^{(2)}_{\bm{q}}(\xi_q^{2},0)]\rangle + c.c.\}
\end{split}\end{equation}
For $p=q$, $\Phi^{pp}_*=0$ and Eq.~\eqref{eq:retcf} is the well-known result \cite{Wen91,WenIJMP92} that the tunneling current to lowest order is given by the imaginary part of the retarded Green function of the tunneling operator for the constriction, evaluated at frequency ${\tilde e} U$.

Let
\begin{equation}
  G^{(k)}_{\bm q}(\xi, s) = \langle V^{(k)}_{\bm{q}}(\xi, s) V^{(k)}_{-\bm{q}}(0,0)\rangle
\end{equation}
be the two-point function of the bare vertex operators $V_{\bm q}$ on the edge component $(k)$. Using translational invariance, $[G_{\bm q}^{(k)}(\xi,s)]^* = G_{\bm q}^{(k)}(-\xi,-s)$, we rewrite the current as
\begin{equation}\begin{split}\label{eq:finalcurrent}
  \langle I\rangle = &\frac{{\tilde e}}{(2\pi)^{2}}\, Re \sum_{p q} t_p t_q^*\\&\int_0^{\infty} ds\, e^{2\pi i\, [{\tilde e} U s - \Phi^{pq}_*] }\, \{ G^{(1)}_{\bm q}(\xi^1_{pq},s) G^{(2)}_{\bm q}(\xi^2_{pq},s) - [G^{(1)}_{\bm q}(\xi^1_{pq}, s) G^{(2)}_{\bm q}(\xi^2_{pq}, s)]^*\}
\end{split}\end{equation}
with $\xi^k_{pq} = \xi^k_p - \xi^k_q$.

The two-point function at finite temperature was given in \eqref{eq:barecorrelator}:
\begin{equation}\begin{split}\label{eq:vertexcfhigh}
  G_{\bm q}(\xi, s) &=
  %\langle V_{\bm{q}}(\xi, s) V_{-\bm{q}}(0,0)\rangle =
  \prod_n \left[ \frac{ \pi T \, i \epsilon }{u_n \sinh \{ \pi T (s + \chi_n u_n^{-1} \xi) \}} \right]^{q_n^2}\\
  &= \prod_n \left[ \frac{ \pi T\, \epsilon }{u_n \sinh \pi T |s + \tau_n|} \right]^{q_n^2} \exp\{i\frac{\pi}{2}\, q_n^2\, \text{sgn}(s + \tau_n) \} \,,
\end{split}\end{equation}
where we use the notation $\tau_n = \chi_n u_n^{-1}\xi$. In the following, we omit the unimportant factor $\prod_n [ \epsilon u_n^{-1} ]^{q_n^2}$ in the two-point function.

\subsection{Tunneling current through a single constriction}

For tunneling through a single constriction ($p=q$), we find from \eqref{eq:finalcurrent} and \eqref{eq:vertexcfhigh} that
\begin{equation}\begin{split}\label{eq:currentintegral0}
  \langle I\rangle = \frac{{\tilde e}}{(2\pi)^{2}}\,
  %(\prod_n [u_n^{(1)}u_n^{(2)}]^{-q_n^2})\,
  (\pi T)^{2g}\, Re\{ 2 i \sin (\pi g) \int_0^{\infty} ds\, e^{2\pi i\, {\tilde e} U s }\, |\sinh \pi T s|^{-2g} \}\,.
\end{split}\end{equation}
In \eqref{eq:currentintegral0}, $g = \sum_n q_n^2$ is the scaling dimension and ${\tilde e} = Q_{el}({\bm q})$ is the electric charge of the tunneling quasi-particle; $U$ is the bias voltage across the constriction. After some algebra (see \ref{app:currentintegrals}), one finds that
\begin{equation}\begin{split}\label{eq:currentintegral1}
  \langle I\rangle = \frac{{\tilde e}}{(2\pi)^{2}}\,
  %(\prod_n [u_n^{(1)}u_n^{(2)}]^{-q_n^2})\,
  (2\pi T)^{2g-1} B(g-i\frac{{\tilde e} U}{T}, g + i\frac{{\tilde e} U}{T} ) \sinh(\pi\frac{{\tilde e} U}{T})\,,
\end{split}\end{equation}
where $B(a, b)$ is the Euler beta function.

At small bias, ${\tilde e} U \ll T$, the tunneling current follows Ohms law,
\begin{equation}\label{eq:directhighT}
  \langle I \rangle \propto U\, T^{2g-2}\, .
\end{equation}
For large bias, ${\tilde e} U \gg T$, we find that
\begin{equation}\label{eq:directlowT}
  \langle I \rangle \propto U\, |U|^{2g-2}\, .
\end{equation}

Note that the ``direct'' tunneling current through a constriction, \eqref{eq:currentintegral0}, does not depend on the charge sector of the initial state $\langle(\cdot)\rangle$, nor does it depend on the propagation speeds $u_n$ or -directions $\chi_n$ of the edge modes. Therefore, measurements of the tunneling current through a constriction may be used to determine the scaling dimension $g$ and the electric charge ${\tilde e}$ of tunneling quasi-particles in quantum Hall effects (see also \cite{Wen91}; experimental fits to this formula were performed in \cite{RaduMarcusPfeifferWest08}).

\subsection{Interference current}\label{sec:interferencecurrent}

Next, we discuss the {\it interference current} resulting from tunneling at {\it two} constrictions. Using \eqref{eq:finalcurrent}, the current is given by
\begin{equation}\begin{split}\label{eq:currentintegral2a}
  \langle I\rangle = &\frac{{\tilde e}}{(2\pi)^{2}}\,
  %(\prod_m [u_m^{(1)}u_m^{(2)}]^{-q_m^2})\,
  (\pi T)^{2g} \sum_{p q} Re\{ 2 i\, e^{- 2\pi i\, \Phi^{pq}_*}\\
   &\int_0^{\infty} ds\, e^{2\pi i\, {\tilde e} U s }\, \sin [\pi \sum_m q_m^2\, f(\tau_{pq,m}^{(1)}, \tau_{pq,m}^{(2)}, -s)]\\
   &\prod_n [\sinh ( \pi T |\tau_{pq,n}^{(1)}+s|) \sinh (\pi T |\tau_{pq,n}^{(2)}+s|)]^{-q_n^2}\}
\end{split}\end{equation}
with
\begin{equation}
  f(\tau_1, \tau_2, s) = \frac{1}{2}[\text{sgn}(\tau_1-s)+\text{sgn}(\tau_2-s)]\, .
\end{equation}
We now perform the sum over $p, q\in\{1,2\}$ and define
\begin{equation}\label{eq:phigamma}
  \Phi^\gamma_* = \Phi^{21}_*\,,
\end{equation}
$\tau_n^{(k)} = \tau_{21,n}^{(k)}$, and $\gamma = \gamma_{21}$. This allows us to display the fact that the tunneling current \eqref{eq:currentintegral2a} {\it does not depend on the initial time} :
\begin{equation}\begin{split}\label{eq:currentintegral2b}
  \langle I\rangle = &\frac{{\tilde e}}{(2\pi)^{2}}\,
  %\, (\prod_m [u_m^{(1)}u_m^{(2)}]^{-q_m^2})\,
  (\pi T)^{2g} \text{Im}\{ 2\, e^{-2 \pi i\, \Phi^\gamma_*}\\
  &\int_{-\infty}^{\infty} ds\, e^{-2\pi i\, {\tilde e} U s }\, \sin [\pi \sum_m q_m^2\, f(\tau_{m}^{(1)}, \tau_{m}^{(2)}, s)]\\
  &\prod_n [\sinh (\pi T |\tau_{n}^{(1)}-s|) \sinh ( \pi T |\tau_{n}^{(2)}-s|)]^{-q_n^2} \}\, .
\end{split}\end{equation}

In contrast to the direct tunneling current \eqref{eq:currentintegral0}, the interference current \eqref{eq:currentintegral2b} depends on the charge sector of the initial state $\langle(\cdot)\rangle$ through the phase $\Phi^\gamma_*$. Before turning to an explicit evaluation of the integral in \eqref{eq:currentintegral2b}, we discuss the effect of charge averaging on the two boundary components involved in the tunneling process that was announced in Sect.~\ref{sec:initialstate}.

\subsubsection{Interference phase, edge charge averaging, and the geometry of the interferometer}

The dependence of the interference phase $\Phi_*^\gamma$, \eqref{eq:phigamma}, on the {\it localized bulk} charges is straightforward to determine. From \eqref{eq:interferencephase3} it is given by the total localized bulk quasi-particle charge {\it inside the interference contour} $\gamma$,
\begin{equation}\begin{split}\label{eq:interferencephasestatic}
  \Phi^\gamma_*(\{{\bm q}^{(k)}\}; \{({\bm q},z)^m\})  = \Phi^\gamma_*(\{{\bm q}^{(k)}\}) + \sum_{z^m\in \text{int} \gamma} {\bm q}\cdot {\bm q}^m\,.\\
\end{split}\end{equation}
The remaining dependence of $\Phi_*^\gamma$ on the {\it edge charges} is more complicated. However, one can derive the following general property: In Sect.~\ref{sec:chargezero}, we have shown that the edge charges can be taken into account by threading additional {\it virtual} $\Af_n$-fluxes through the holes of the sample. Let $\partial^{(0)}\!\Omega$ be the outer boundary component  and let $\{\partial^{(k)}\!\Omega\}_{k>0}$ be the boundary components enclosing holes in the sample (see Fig.~\ref{fig:sample}). Then, using \eqref{eq:interferencephase3}, we find that
\begin{equation}\begin{split}\label{eq:phiproperty}
  \Phi^\gamma_*({\bm q}^{(0)} + &\sum_{k>0} \delta {\bm q}^{(k)}, \{{\bm q}^{(k)} - \delta {\bm q}^{(k)}\}_{k>0})\\
  &= \Phi^\gamma_*({\bm q}^{(0)}, \{{\bm q}^{(k)}\}_{k>0}) + \sum_{\substack{k>0,\\\partial^{(k)}\!\Omega\, \subseteq\, \text{int}\gamma}} {\bm q}\cdot \delta{\bm q}^{(k)}\, .
\end{split}\end{equation}

As discussed in Sect.~\ref{sec:initialstate}, an appropriate initial state, $\langle(\cdot)\rangle$, for perturbation theory is a superposition of states in the charge superselection sectors of the two boundary components involved in the tunneling process, Eq.~\eqref{eq:delocalizedsector}. The current corresponding to such an initial state can be obtained from the current in a charge sector, Eq.~\eqref{eq:currentintegral2b}, by an appropriate sum over different charge sectors. The only dependence of the interference current \eqref{eq:currentintegral2b} on the sector comes from the interference phase $\Phi^\gamma_*$. Hence, we obtain
\begin{equation}
  \langle I \rangle_{\bm q} \propto \sum_N \langle I \rangle_N \propto \sum_{N} \exp\{ 2\pi i \Phi_*^\gamma({\bm q}^{(0)}, {\bm q}^{(1)} + N {\bm q}, {\bm q}^{(2)} - N {\bm q}, {\bm q}^{(3)}, \ldots ) \}\,,
\end{equation}
where $\langle (\cdot) \rangle_{\bm q}$ is a state in \eqref{eq:delocalizedsector} and $\langle (\cdot) \rangle_N$ is a state in a sector with charges ${\bm q}^{(1)} + N {\bm q}$ and ${\bm q}^{(2)} - N {\bm q}$ on $\partial^{(1)}\!\Omega$ and $\partial^{(2)}\!\Omega$, respectively.
We can now use Eq.~\eqref{eq:phiproperty} to evaluate the interference current for a charge-averaged initial state, $\langle I \rangle_{\bm q}$. We find that it {\it vanishes} if one and only one of the holes corresponding to a boundary component involved in the tunneling process {\it lies inside the interference contour} $\gamma$.\footnote{This statement is insensitive to whether edge tunneling occurs from the outer boundary component $\partial^{(0)}\!\Omega$, or if it occurs between two boundary components enclosing holes in the sample. We only discuss the latter case here, but the former situation is similar.} This follows from the fact that the quantity ${\bm q}\cdot{\bm q} = \sum_n \chi_n q_n^2\in \mathbb{Q}\setminus\mathbb{N}$ is fractional, and, therefore, 
\begin{equation}
  \langle I \rangle_{\bm q} \propto \sum_{N} [\exp \{2\pi i {\bm q}\cdot{\bm q}\}]^N = 0\, .
\end{equation}

\begin{figure}[!h]
\center
\includegraphics[width=0.7\textwidth]{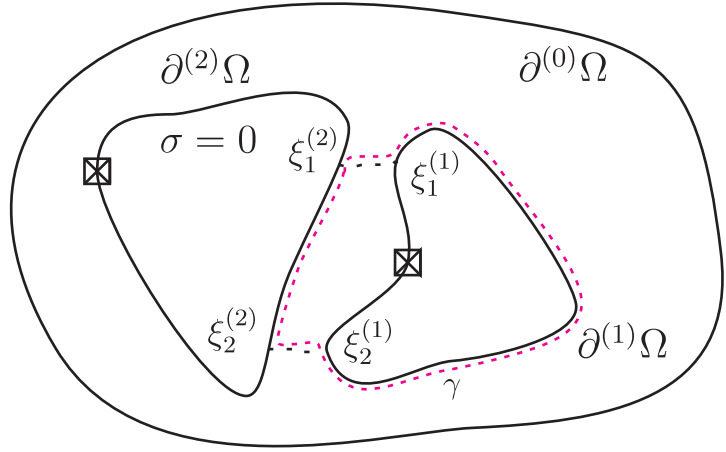}
\caption{An interference contour $\gamma$ for the geometry of a Mach-Zehnder interferometer (red dashed line). The Ohmic contacts on the two boundary components $\partial^{(1)}\!\Omega$ and $\partial^{(2)}\!\Omega$ involved in the tunneling process are denoted by squares with a cross. Fractional charge averaging on the two boundary components leads to the absence of an interference term in the inter-edge tunneling current.
\label{fig:MZ1}
}
\end{figure}

\begin{figure}[!h]
\center
\includegraphics[width=0.7\textwidth]{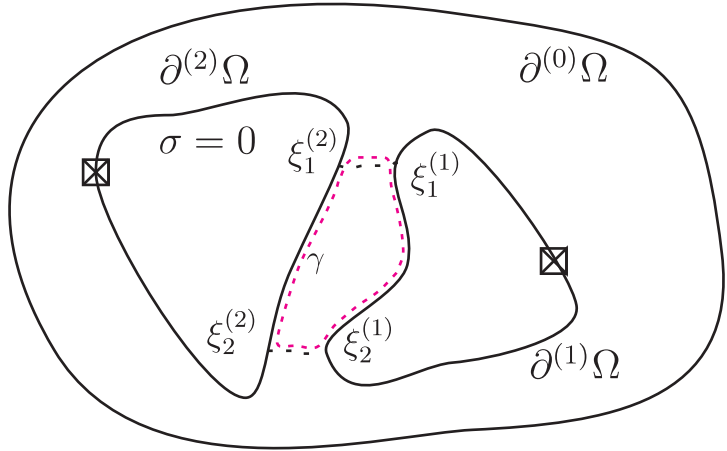}
\caption{An interference contour $\gamma$ for the geometry of a Fabry-Perot interferometer (red dashed line). Charge averaging does {\it not} affect the inter-edge tunneling current and an interference term is present.
\label{fig:FP1}
}
\end{figure}

In the terminology of \cite{LevkivskyiBoyarskyFrohlichSukhorukov09}, the case when exactly one hole corresponding to a tunneling boundary component lies inside the interference contour is called a {\it Mach-Zehnder} interferometer; (Fig.~\ref{fig:MZ1}). The other case, when no hole or two holes lie inside the interference contour, is called a {\it Fabry-Perot} interferometer; (Fig~\ref{fig:FP1}). Fractional charge delocalization on the two tunneling boundary components in the fractional quantum Hall effect results in a cancelation of the interference term in the inter-edge tunneling current for the geometry of a Mach-Zehnder interferometer. In the case of the Fabry-Perot interferometer, however, the interference term is not affected by the charge delocalization. This is a good place to remind the reader that the geometry of the interference contour $\gamma$ is specified by the location of the Ohmic contacts on the two tunneling boundary components.
In the following, we assume the case of an interference path with the geometry of a Fabry-Perot interferometer.

\subsubsection{Chiral edge with a single propagation speed}

As a first simplification, let us consider the interference current in the case when all modes on each edge involved in the tunneling have the same chirality and propagation speed, i.e., $\chi_n=\chi$ and $u_n^{(k)}=u^{(k)}$ for all $n = 1,\ldots, N$. In this case, the current \eqref{eq:currentintegral2b} is given by
\begin{equation}\begin{split}\label{eq:currentintegral3}
  \langle I\rangle = &\frac{{\tilde e}}{(2\pi)^{2}}\, (\pi T)^{2g}\, 2 \sin (\pi g) \text{Im} \{  e^{ - 2 \pi i\, \Phi_*^\gamma} \int_{-\infty}^{\infty} ds\, e^{- 2\pi i\, \tilde e U s }\, \\
  &[\sinh ( \pi T |\tau^{(1)}-s| ) \sinh ( \pi T |\tau^{(2)}-s|)]^{-g} f(\tau^{(1)}, \tau^{(2)}, s)\}\, .
\end{split}\end{equation}
Clearly, the current changes sign when the bias voltage, $U$, changes sign; in the following, we assume that $U>0$. The integral \eqref{eq:currentintegral3} is evaluated in \ref{app:interferenceintegral}.
The result is
\begin{equation}\begin{split}\label{eq:currentintegral4}
  \langle I\rangle =
  \frac{ {\tilde e}\, (2 \pi T)^{2 g - 1}}{\pi\, \Gamma(g)}\,  &e^{- \pi g T |\tau^{(2)}-\tau^{(1)}|} \cos 2\pi (\Phi_*^\gamma + \frac{\tilde e U}{2}[\tau^{(1)} + \tau^{(2)}])\\ &\text{Im}\{ e^{i\pi \tilde e U |\tau^{(2)} -\tau^{(1)}|} C_g(\tilde e U,|\tau^{(2)} -\tau^{(1)}|, T) \}
\end{split}\end{equation}
with $C_g$ given in terms of the hypergeometric function,
\begin{equation}
  C_g(\mu,\tau, T) = \,_2F_1 (g, g - i\beta\mu; 1-i\beta\mu; e^{ - 2 \pi g T \tau})\, \frac{\Gamma(g - i\beta\mu)}{\Gamma(1 - i\beta\mu)}\, .
\end{equation}

In the limit of low temperatures ($T\ll \tilde e U, |\tau^{(k)}|^{-1}$), Eq.~\eqref{eq:currentintegral4} reduces to
\begin{equation}\begin{split}\label{eq:currentintegral5}
  \langle I\rangle =
  \frac{ {\tilde e}\, \sqrt{\pi} }{2\pi\, \Gamma(g)}\, &\cos 2\pi(\Phi_*^\gamma + \frac{\tilde e U}{2}[\tau^{(1)}+\tau^{(2)}] )\\
  &\left(\frac{2\pi \tilde e U}{|\tau^{(2)}-\tau^{(1)}|}\right)^{g-\frac{1}{2}} J_{g-\frac{1}{2}}(\pi\tilde e U |\tau^{(2)}-\tau^{(1)}| )\, ,
\end{split}\end{equation}
where $J_{\nu}(x)$ is the Bessel function of index $\nu$. In the limit of high temperatures (for $T\gg \tilde e U,|\tau^{(k)}|^{-1}$), we find
\begin{equation}\begin{split}\label{eq:highTlimitcurrent1}
  \langle I\rangle \propto\,\;
  &T^{2 g - 1}\,  e^{- \pi g T |\tau^{(2)}-\tau^{(1)}|}\\
  &\cos 2\pi (\Phi_*^\gamma + \frac{\tilde e U}{2}[\tau^{(1)} + \tau^{(2)}])\,
  \sin(\pi \tilde e U |\tau^{(2)} -\tau^{(1)}|)\, .
\end{split}\end{equation}

From Eqs.~\eqref{eq:currentintegral5} and \eqref{eq:highTlimitcurrent1}, we see that our weak tunneling theory predicts an interference current that oscillates as a function of voltage, with periods $\Delta U = 2 /({\tilde e}|\tau^{(2)} \pm \tau^{(1)}|)$. At large voltage and low temperatures, the envelope of the oscillations behaves like $|\mu|^{g-1}$. At high temperature, the envelope follows $T^{2g-1} e^{-\pi g T |\tau^{(2)}-\tau^{(1)}|}$ and the interference current is exponentially suppressed.

In the geometry of the Fabry-Perot interferometer (Fig.~\ref{fig:FP1}), we have $\tau^{(1)}\tau^{(2)}<0$.
For a symmetric interferometer  with $\tau := |\tau^{(2)}| \simeq |\tau^{(1)}|$, the oscillations of the interference current as a function of voltage (for $T \ll \tilde e U,\tau^{-1}$) stem from the Bessel function,
\begin{equation}\label{eq:currentintegral6}
    \langle I\rangle =
  \frac{ {\tilde e}\,\sqrt{\pi} }{2\pi\, \Gamma(g)}\, \cos (2\pi\Phi_*^\gamma)\, \left(\frac{2\pi \tilde e U}{\tau}\right)^{g-\frac{1}{2}} J_{g-\frac{1}{2}}(2 \pi \tilde e U \tau)\,.
\end{equation}
For $\tilde e U \tau \ll 1$, we find that $\langle I\rangle \propto |\tilde e U|^{2g-1}$, and we recover the result \eqref{eq:directlowT} for the tunneling current through a single constriction. In the limit $\tilde e U \tau \gg 1$, we have $\langle I\rangle \propto |\tilde e U|^{g-1}$. It is interesting to observe that the amplitudes of interference current oscillations as a function of voltage change their power law at the scale $\tilde e U \sim \tau^{-1}$.

\subsubsection{Interferometer with two distinct propagation speeds}

Next, we consider a quantum Hall edge exhibiting (two families of) gapless excitations that propagate at two distinct propagation speeds, $u_1$ and $u_2$, with $u_1 \ll u_2$. The two propagation speeds are identical on both tunneling boundary components $\partial^{(1)}\!\Omega$ and $\partial^{(2)}\!\Omega$. In fact, it is believed that electrically charged modes ($Q_n\neq 0$) propagate at a significantly higher speed than neutral modes \cite{LopezFradkin99,LeeWen91,OverboschChamon09}. As before, we suppose that the chiralities of all modes are identical, i.e., $\chi_n=1$ for all $n$. The scaling dimension of the two modes is denoted by $g_1$ and $g_2$, and we let $g=g_1+g_2$.

The general expression for the interference current is as given in \eqref{eq:currentintegral2b}. In the limit of a {\it symmetric interferometer}, $|\xi^{(1)}| \simeq |\xi^{(2)}|$, we let $\tau = \tau^{(1)}_1 \simeq - \tau^{(2)}_1 > 0$. In this case, the interference current is given by
\begin{equation}\begin{split}\label{eq:currentsymmetric}
  \langle I \rangle = 2\cos(2\pi \Phi_*^\gamma )\, \text{Im}\,\{  \sin(\pi g)\, e^{2\pi i \tilde e U \tau} K_2(2 \tau ,\tau) + \sin(\pi g_2) \hat K_2(\tau,-\tau) \}\,.
\end{split}\end{equation}
The functions $K_2(x,y)$ and $\hat K_2(x,y)$ are displayed in \ref{app:2velocities} in terms of Appell's hypergeometric function of two variables, $F_1$; [Eqs.~\eqref{eq:K2eval} and \eqref{eq:K2hateval}].

In the limit of an {\it asymmetric interferometer}, with $\xi^{(1)} \ll \xi^{(2)}$, we let $\tau = \tau^{(1)}_1>0$, and $\tau \gg |\tau^{(2)}_1|$. In the limit of low temperatures ($T\ll \tilde e U, \tau^{-1}$), the interference current is then given by
\begin{equation}\begin{split}\label{eq:currentasymmetric}
  \langle I \rangle = \frac{{\tilde e}}{2\pi}\tau^{1-2g}\, \text{Im} \{ \sin (\pi g) &[e^{2\pi i\, \Phi_*^\gamma} e^{2\pi i\, \tilde e U \tau} \tilde U(1-g_1,2-2g,-2\pi i\, \tilde e U \tau)\\
  &+ e^{-2\pi i\Phi_*^\gamma} \tilde U(1-g-g_2,2-2g,-2\pi i\, \tilde e U \tau)]\\
  + \sin (\pi g_2) &\tilde M(1-g-g_2, 2-2g, 2\pi i\, \tilde e U \tau) \}\, .
\end{split}\end{equation}
The functions $\tilde M$ and $\tilde U$ are regularized versions of Kummer's confluent hypergeometric functions $M$ (also denoted by $_1F_1$) and $U$ (Tricomi function); see Eqs.~\eqref{eq:Mtilde} and \eqref{eq:Utilde} in \ref{app:2velocities}. We are not able to obtain a similar simple expression of the current in the low-temperature limit for a {\it symmetric} interferometer; Eq.~\eqref{eq:currentsymmetric} has to be used in that case.

Setting $g_2=0$ and $g_1=g$, the expressions \eqref{eq:currentsymmetric} and \eqref{eq:currentasymmetric} reduce to the interference current for a single propagation speed, Eq.~\eqref{eq:currentintegral4}.

\subsection{Interference current for dynamical modifications of the sample}\label{sec:dynmod}

Next, we want to discuss the possible effects on the tunneling current of {\it dynamical}, i.e., time-dependent changes applied to the quantum Hall sample. More precisely, we are interested in slowly time-dependent modifications that are switched on {\it after the tunneling current has equilibrated}, at a time $t>t_{eql}$.

\subsubsection{Increase of magnetic flux}

Here, we discuss the effect on the tunneling current when the magnetic flux through the sample is increased after the equilibration time $t_{eql}$. The electromagnetic gauge fields enter the tunneling current \eqref{eq:current2} through the interference phase \eqref{eq:interferencephase}.

Let us first consider the case of a {\it static} magnetic flux through a {\it hole} in the incompressible fluid. Suppose that the interference contour $\gamma$ winds around that hole. In a static situation, the magnetic flux through the hole, $\Phi^{\text{em}}_{\text{hole}}$, appears on the right-hand side of Eq.~\eqref{eq:interferencephase2}, and gives a contribution $\tilde e\Phi^{\text{em}}_{\text{hole}}$ to $\Phi_*^\gamma$. One concludes that the interference current, \eqref{eq:currentintegral6} or \eqref{eq:currentsymmetric}, is periodic in $\Phi^{\text{em}}_{\text{hole}}$ with an ``Aharonov-Bohm'' period $1/{\tilde e}>1$. This is in contradiction to the well-known fact that one unit of magnetic flux ($\delta\Phi^{em}=1$ in our units) through a hole in an electronic system can always be removed by a gauge transformation \cite{ByersYang61}. In a microscopic approach, resolutions to this apparent paradox have been proposed \cite{Lee90,ThoulessGefer91,Kane03,LFSprog}; and can be summarized by the following scenario: When  magnetic flux through a hole is adiabatically increased by one unit, the system ends up in an excited state. As the system relaxes to its ground state, a fractional quasi-particle is transferred between the boundary components which restores the electronic period $\delta\Phi^{em}$.

Let us analyze a process of dynamic increase of magnetic flux, using the low-energy field theory approach of this paper. It is clear that the static expression \eqref{eq:interferencephase2} for the interference phase $\Phi^{pq}(t',t)$ cannot be used in this situation, for $t>t_{eql}$. Instead, a possibly time-dependent term is added to the phase $\Phi_*^\gamma$ in the interference current displayed in Sect.~\ref{sec:interferencecurrent},
\begin{equation}\label{eq:phasechange}
  \Phi_*^\gamma \mapsto \Phi_*^\gamma + \tilde\Phi^\gamma(t)\, .
\end{equation}
For simplicity of notation, let us set the speeds $u_n=1$ and chiralities $\chi_n=1$ in this section. From \eqref{eq:interferencephase}, neglecting the line integrals of the bulk fields $\Af_n$ for sufficiently narrow constrictions, the additional phase $\tilde\Phi^\gamma(t)$ in \eqref{eq:phasechange} is given by
\begin{equation}\begin{split}\label{eq:phasevariation}
\tilde\Phi^\gamma(t) &= \sum_n \chi_n q_n \int_{\gamma}\partial_1 p_n[\tilde\af_n](\xi,t)\, d\xi\\
  &= \sum_n \chi_n q_n \{ p^{(1)}_n(\xi^1_2,t) - p^{(1)}_n(\xi^1_1,t) + p^{(2)}_n(\xi^2_1,t) - p^{(2)}_n(\xi^2_2,t) \}\, .
\end{split}\end{equation}
The fields $\tilde \af_n$ in \eqref{eq:phasevariation} are the slowly time-dependent external gauge fields switched on for $t>t_{eql}$, {\it in addition to the static gauge fields}. Note that $p^{(k)}_n[\tilde\af_n](\xi,t)$ are single-valued functions on $\partial^{(k)}\!\Omega$, for all times.

We calculate the phase \eqref{eq:phasevariation} for the special case of an increase of magnetic flux through surfaces bounded by $\partial^{(k)}\!\Omega$. For this purpose, we consider electromagnetic gauge fields of the form
\begin{equation}\label{eq:fluxform}
  \tilde a^{(k)}_1(\xi,t) = f^{(k)}(\xi) \Phi^{(k)}_{em}(t)
\end{equation}
and $\tilde a^{(k)}_0=0$, where $f^{(k)}(\xi)$ are continuous periodic functions on $\partial^{(k)}\!\Omega$ with $\int_0^{L^{(k)}} d\xi\, f^{(k)}(\xi) = 1$ and $L^{(k)} = |\partial^{(k)}\!\Omega|$; $\Phi^{(k)}_{em}(t)$ is the magnetic flux through a surface bounded by $\partial^{(k)}\!\Omega$, which increases in time from $\Phi^{(k)}_{em}(t\leq t_{eql})=0$ to $\Phi^{(k)}_{em}(t\gg t_{eql})=\Phi^{(k)\infty}_{em}$. By Eq.~\eqref{eq:ourp}, the functions $p_n[\tilde\af_n]$ are given by
\begin{equation}
  p^{(k)}_n[\tilde\af_n](\xi,t) = Q_n \int^t ds\, f^{(k)}(\xi + t-s) \Phi^{(k)}_{em}(s)\, .
\end{equation}
It is straightforward to see that, for $t\gg t_{eql}$,
\begin{equation}\begin{split}
\label{eq:shiftintegral}
  p^{(k)}_n&[\tilde\af_n](\xi_2,t) - p^{(k)}_n[\tilde\af_n](\xi_1,t)\\
  &= Q_n \int^t ds\, \{f^{(k)}(\xi_2 + t-s) - f^{(k)}(\xi_1 + t-s) \} \Phi^{(k)}_{em}(s)\\
  &\simeq \frac{\xi_2-\xi_1}{L^{(k)}}\, Q_n \Phi^{(k)\infty}_{em} + \ldots\,,
\end{split}\end{equation}
where ``$\ldots$'' stands for terms that are highly oscillatory in time, with frequency $\omega\gtrsim u/L^{(k)}$; $u$ is the propagation speed of the slowest edge mode, and $L^{(k)}$ is the length of the boundary component. We expect that such high-frequency oscillations in the tunneling current are unobservable in transport measurements.\footnote{In fact, the high-frequency oscillations in \eqref{eq:shiftintegral} stem from the propagation of variations in the edge charge distribution around the biased edge components. It is plausible that these oscillations are completely suppressed by the Ohmic contact.} The first term on the right-hand side of \eqref{eq:shiftintegral}, however, is negligible by our hypothesis \eqref{eq:condition}: The distance between neighboring tunneling constrictions is much smaller than the total length of the edge. If this were not the case, the two-point functions used in this calculation would be strongly affected by the Ohmic contacts and our calculation is invalid. Therefore, in the case under consideration, the time-dependent phase shift \eqref{eq:phasevariation} in \eqref{eq:phasechange} is unobservable; $\Phi^\gamma(t)\simeq 0$ for an arbitrary variation of magnetic flux through the holes of the sample.

Note that the form \eqref{eq:fluxform} of the electromagnetic field is not only suitable for an insertion of magnetic flux through the boundary components $\partial^{(k)}\!\Omega$ involved in the tunneling process. The same argument can be used for the process of a dynamical increase of magnetic flux at an {\it arbitrary location} of the sample,\footnote{Of course, the change of magnetic flux must be sufficiently small such that the incompressible state remains intact.} by using functions $f^{(k)}(\xi)$ in \eqref{eq:fluxform} such that $\int_0^{L^{(k)}} d\xi\, f^{(k)}(\xi) = 0$. This only gives highly oscillatory, unobservable contributions to the phase \eqref{eq:phasevariation}.

\subsubsection{Deformation of the interference contour}

In the previous section we concluded that the interference term in the tunneling current remains constant under arbitrary adiabatic variations of the magnetic flux through the sample. However, a way to observe ``Aharonov-Bohm'' type effects is to {\it deform the interference contour}, $\gamma\mapsto\gamma(t)$. To do so, the boundary components $\partial^{(k)}\!\Omega$ involved in the tunneling process need to be deformed by some external force (e.g., by a gate voltage). Assuming that the distance between the tunneling constrictions, $|\xi^k_2-\xi^k_1|$, remains constant, the interference term in the tunneling current acquires a time dependence due to the phase
\begin{equation}\label{eq:phase3}
  \Phi_*^\gamma(t) = \Phi_*^{\gamma(t)} = \tilde e \Phi_{\gamma(t)}^{\text{\text{em}}}\,.% + \sum_n \chi_n q_n \Phi_{\gamma(t)n}\,,
\end{equation}
where $\Phi_{\gamma(t)}^{\text{\text{em}}}$ is the magnetic flux through the interference contour at time $t$, $\gamma(t)$. Hence, we expect an ``Aharonov-Bohm'' period $1/\tilde e > 1$ in the magnetic flux through the interference contour when the flux is varied by deformation of the boundary components. Equation \eqref{eq:phase3} provides access to the fractional charge $\tilde e$ of the quasi-particles tunneling at the constrictions. Note that, in \eqref{eq:phase3}, we assume that the localized bulk quasi-particles remain far from the boundary and from the tunneling constrictions. The process when a bulk quasi-particle is dragged across the interference contour is discussed next.

\subsubsection{Dragging around localized bulk quasi-particles}\label{sec:dragging}

\begin{figure}[!h]
\center
\includegraphics[width=0.6\textwidth]{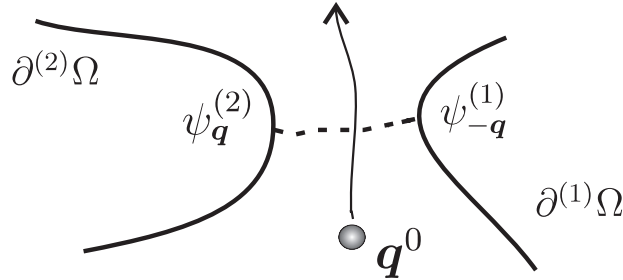}
\caption{Localized bulk quasi-particle with charge ${\bm q}^0$ is moved across a tunneling constriction.
\label{fig:qpconstriction}
}
\end{figure}

Next, we analyze the effect of dragging a localized bulk quasi-particle around the sample by some external agent. As long as the quasi-particle is located far from the edge and far from the constrictions, formulae similar to \eqref{eq:fluxform} and \eqref{eq:shiftintegral} can be used to show that the interference phase $\Phi_*^\gamma$ remains constant. However, the situation is different when a quasi-particle is moved across the interference contour $\gamma$ at a constriction (see Fig.~\ref{fig:qpconstriction}).
In this case, the line integral of the bulk field $\Af_n$ in the tunneling operator \eqref{eq:Top} cannot be neglected. When a bulk quasi-particle with charge ${\bm q}^0$ is moved across the constriction, it follows from \eqref{eq:interferencephase} that the interference phase $\Phi_*^\gamma$ changes by the amount
\begin{equation}\label{eq:qpconstriction}
  \Delta\Phi_*^\gamma = {\bm q}^0\cdot {\bm q} = \sum_n \chi_n q_n^0 q_n\,,
\end{equation}
where ${\bm q}$ is the charge vector of the tunneling quasi-particle. This is the same as the static result, \eqref{eq:interferencephasestatic}.

Equation \eqref{eq:qpconstriction} provides experimental access to the projection of the tunneling quasi-particle charge vector ${\bm q}$ to a charge vector of a localized bulk quasi-particle ${\bm q}^0$. In Eq.~\eqref{eq:phase3}, note that only the electrical charge modes (with $Q_n\neq 0$) contribute to the quasi-particle charge ${\tilde e}$. The phase jumps \eqref{eq:qpconstriction} due to bulk quasi-particles leaving the interference contour, however, probe the neutral modes, too.

It is tempting to expect a similar result for the situation when a localized bulk charge is brought close to the edge and is then absorbed by the edge. Total charge conservation implies that the edge charge is increased by the amount of absorbed bulk charge. However, the effect of such a process on the interference phase $\Phi_*^\gamma$ is unclear, since we do not know its explicit dependence on the edge charges in \eqref{eq:interferencephasestatic}.

\subsection{Short digression on energy scales}

In the following, we give some physically relevant scales which should be used to describe realistic experimental samples. Transport experiments in quantum Hall samples are done (see \cite{RaduMarcusPfeifferWest08}) in a temperature range of $T \sim 10 - 30 \, m\!K$. The applied bias voltage at the constrictions is $U \lesssim 100\, \mu\!V$, corresponding to $e U/k_B \lesssim 1\, K$ ($1 \mu eV\hat = 11.6\, mK$). The linear size of the quantum Hall edge is $L\sim 5\cdot 10^{-3}\, m$.
%$\mu/( 2 \pi ) e/{\tilde e} = $
In a typical interference experiment, the distance between tunneling contacts is of the order of $\xi \sim 2\cdot10^{-3}\, m$.

It is expected on general grounds \cite{LopezFradkin99,LeeWen91,OverboschChamon09} that the propagation speed of the charge modes, $u_c$, is essentially given by a scale set by the the Coulomb interaction energy, $u_c \simeq e^2/(\varepsilon\hbar)$. For a realistic value of the dielectric constant, $\varepsilon\simeq 10$, one gets the estimate $u_c \simeq 10^{5}\, m s^{-1}$. In contrast, the propagation speed of the neutral modes, $u_n$, is expected to be much smaller.

Using this speed estimate for the charge mode, the energy scale corresponding to the finite length of the edge is given by $T_L = \hbar\, u_c / (k_B L) \sim 1\, m\!K$. We see that, in order to describe current experiments, we are in the range $T_L \ll T$ and $L$ can effectively be taken to infinity. Moreover, the bias voltage is of the order of the temperature, $e U /(2\pi) \sim T$; Hence, it should be possible, with currently available techniques, to probe both low- and high-temperature limits of the tunneling current discussed in this paper.

The distance between tunneling constrictions, $\xi$, corresponds to a temperature scale $T_\xi \sim 2\, m\!K$. This is the smallest remaining energy scale in the problem, i.e., $T_{\xi} \ll T, eU$. Our calculation predicts oscillations of the interference current as a function of bias voltage $U$ with periods $\propto |\xi^1\pm\xi^2|^{-1}$, see Eqs.~\eqref{eq:currentintegral5} and \eqref{eq:highTlimitcurrent1}. In present day experiments, $|\xi^1 + \xi^2|^{-1}$ leads therefore to high-frequency oscillations which we expect to be unobservable. However, for a sufficiently symmetric interferometer, one may hope to observe oscillations with period $\propto |\xi^1 - \xi^2|^{-1}$. From such measurements, it may even be possible to extract the propagation speed of the charge mode.

\section*{Acknowledgments}
We thank A.~Boyarsky, V.~Cheianov, I.~Levkivskyi, and E.~Sukhorukov for very useful discussions. This work has profited from insights due to earlier collaborations of the senior Author \cite{LFSprog,LevkivskyiBoyarskyFrohlichSukhorukov09}, especially for the part concerning quasi-particle interferometry. This work was supported by the Swiss National Science Foundation.

\appendix

\def\thesubsection       {\thesection.\arabic{section}}

\section{Conserved current \label{app:conscurr}}

In this appendix we show that the conserved current on the edge, $\jeoh^\mu_n$, Eq.~\eqref{eq:conscurrents}, can be written as in \eqref{eq:conscurrents2}. Furthermore, we derive the solutions \eqref{eq:ourp} and \eqref{eq:ourp0} for the function $p_n[\af_n]$ in \eqref{eq:defvarphi}. To keep our notation compact, we omit the mode index $n$ and write $a$, instead of $\af_n$, for the gauge field. For simplicity, we set the propagation speed $u = 1$ and the chirality $\chi = 1$.

The conserved current is defined as
\begin{equation}\label{eq:cc1}
  \jeoh^\mu = \frac{1}{2}(g^{\mu\nu} + \epsilon^{\mu\nu})(\partial_\nu \varphi + \partial_\nu p + a_{\nu}) - \epsilon^{\mu\nu} a_{\nu}\,,
\end{equation}
with
\begin{equation}
  \phi(\xi,t) = \varphi(\xi,t) + p[a](\xi,t)\, .
\end{equation}
The function $p[a](\xi,t)$, for $\xi\in\partial\Omega$, is a solution of the equation \eqref{eq:motion3}, i.e.,
\begin{equation}\label{eq:cc2}
  \Box p = (\epsilon^{\mu\nu} - g^{\mu\nu})\partial_{\mu} a_{\nu}\, .
\end{equation}

We introduce light-cone coordinates $\xi^{\pm} = \xi^0\pm \xi^1$, and $\partial_{\pm} = (\partial_0\pm \partial_1)/2$, and we define $a_\pm = (a_0\pm a_1)/2$. In terms of light-cone coordinates, Eqs.~\eqref{eq:cc1} and \eqref{eq:cc2} can be written as
\begin{equation}\label{eq:currentLC}%LC meaning Light Cone
  *\jeoh = \epsilon_{\mu\nu} \jeoh^\mu d\xi^\nu = (\partial_+ \varphi + \partial_+ p + a_+)\, d\xi^+ - a_\mu d\xi^\mu = \partial_+ \varphi\, d\xi^+ - (\partial_- p + a_-)\, d\xi^- + dp
\end{equation}
and
\begin{equation}\label{eq:motionLC}
  \partial_+(\partial_- p + a_- ) = 0\, .
\end{equation}

The solution to \eqref{eq:motionLC} with Cauchy data
\begin{equation}\begin{split}\label{eq:cauchy}
  p(\xi, t=0) &= p_0(\xi)\,,\\
  \partial_0 p(\xi, t=0) &= \dot p_0(\xi)\,
\end{split}\end{equation}
at the time $t=0$, is given by
\begin{equation}\begin{split}\label{eq:psolution0}
  p(\xi,t) = &\frac{1}{2}[p_0(\xi+t) + p_0(\xi-t)] + \int_{0}^{t} ds\, \dot p_0(\xi + t - 2 s)\\
  & + 2\int_{0}^{t} ds\, [a_-(\xi+t-2s, 0) - a_-(\xi+t-s, s)]\, .
\end{split}
\end{equation}

Let us compute the function $\partial_- p + a_-$ appearing in \eqref{eq:currentLC} and using \eqref{eq:psolution0}. We have that
\begin{equation}
  \partial_- p(\xi,t) + a_-(\xi, t) = \frac{1}{2}[\dot p_0(\xi-t) - \partial_1 p_0(\xi-t)] + a_-(\xi - t, 0)\, .
\end{equation}
We see that, for a suitable choice of Cauchy data, $\partial_- p + a_-$ can be made to vanish for all times. Such a choice is
\begin{equation}\begin{split}
\label{eq:specialcauchy}
  \dot p_0(\xi) &= -a_0(\xi,0)\,,\\
  \partial_1 p_0(\xi) &= -a_1(\xi,0)\, .
\end{split}\end{equation}
In that case, $\partial_- p + a_- = 0$. Plugging this into \eqref{eq:currentLC} yields the result \eqref{eq:conscurrents2}. Furthermore, using \eqref{eq:specialcauchy}, the solution \eqref{eq:psolution0} for $p$ becomes
\begin{equation}\begin{split}\label{eq:psolution1}
  p(\xi,t) = p_0(\xi+t) - 2 \int_{0}^{t} ds\, a_-(\xi+t-s, s)\, ,
\end{split}
\end{equation}
with
\begin{equation}
  p_0(\xi) = - \int_0^\xi d\bar\xi\, a_1(\bar\xi,0)\,,
\end{equation}
which corresponds to \eqref{eq:ourp} and \eqref{eq:ourp0}.

Note that our choice of Cauchy data \eqref{eq:specialcauchy} has a further advantage of respecting the gauge-dependence of $p(\xi,t)$ at the initial time $t=0$. We have that
\begin{equation}
  \partial_\mu p(\xi,t)|_{t=0} = -a_\mu(\xi, 0)\,.
\end{equation}
Under a general gauge transformation, \eqref{eq:psolution1} transforms like
\begin{equation}\begin{split}
  p[a + d\alpha](\xi,t) &= p[a](\xi,t) - \int_0^{\xi+t} d\bar\xi\, \partial_1\alpha(\bar\xi,0) - \int_0^t ds\, \frac{d}{ds} \{ \alpha(\xi+t-s,s) \}\\
  &= p[a](\xi,t) - \alpha(\xi,t) + \alpha(0,0)\, .
\end{split}\end{equation}
Without loss of generality, we require $\alpha(0,0)=0$ to recover the correct gauge dependence of the field $\phi$, Eq.~\eqref{eq:phitildetransf}.

\section{Some special integrals}\label{app:currentintegrals}

In this appendix, we sometimes use $\beta = T^{-1}$ to denote the inverse temperature. $\mu = \tilde e U$ is the chemical potential at bias voltage $U$. To evaluate the direct tunneling current through a single constriction, we need to compute the following integral:
\begin{equation}\label{eq:K0m}\begin{split}
  K_0(T) &= (\pi T)^{2g} \int_0^\infty ds\, [\sinh \pi T s ]^{-2 g} e^{2 \pi i \mu s }\\
   &= (2\pi T)^{2 g}\int_0^\infty \frac{ds\, e^{2 \pi s (i \mu - T g)}}{(1-e^{-2 \pi T s})^{2g}}\, .
\end{split}\end{equation}
Substituting $u = e^{-2 \pi T s}$, we find that
\begin{equation}\label{eq:K0}\begin{split}
  K_0(T) &= (2 \pi T)^{2 g} \int_0^1 \frac{u^{g-i \beta \mu}}{(1-u)^{2 g}} \frac{du}{2\pi T u}\\
   &= (2 \pi T)^{2 g-1} \int_0^1 u^{g-i \beta \mu-1}\, (1-u)^{-2g}\, du\, .
\end{split}\end{equation}
This is of the form of the Euler Beta function given by
\begin{equation}
  B(a,b) = \int_0^1 du\, u^{a-1}(1-u)^{b-1}\, .
\end{equation}
Hence, we find that
\begin{equation}\label{eq:K00}
  K_0(T) = (2 \pi T)^{2 g-1} B(g-i \beta \mu, 1-2g )\, .
\end{equation}
This expression can be further simplified by writing it in terms of Gamma functions,
\begin{equation}
  B(a, b) = \frac{\Gamma(a)\Gamma(b)}{\Gamma(a+b)}
\end{equation}
and using that
\begin{equation}
  \Gamma(x)\Gamma(1-x) = \frac{\pi}{\sin\pi x}\, .
\end{equation}
We then arrive at
\begin{equation}\label{eq:K000}
  K_0(T) = (2 \pi T)^{2 g-1}\, B(g-i\beta \mu, g+i \beta \mu)\, \frac{\sin\pi(g + i \beta \mu)}{\sin 2\pi g}\, .
\end{equation}

The limit $\mu\gg T$ is most easily found from \eqref{eq:K00}: We have $B(a,b)\simeq \Gamma(a)\, b^{-a}$, for $a\gg 1$ and $b$ fixed. Hence,
\begin{equation}\label{eq;K0low}\begin{split}
  K_0(T\ll\mu) &= (2\pi)^{2g-1} \frac{\pi}{\Gamma(2g)\sin 2 \pi g} (-i\mu)^{2g-1}\\
   &= |2\pi\mu|^{2g-1} e^{-i\pi(g-1/2) \text{sgn}(\mu)} \frac{\pi}{\Gamma(2g)\sin 2 \pi g}\, .
\end{split}\end{equation}
The limit $\mu\ll T$ is immediate from \eqref{eq:K000}.

From \eqref{eq:currentintegral0}, the direct tunneling current through a single constriction is therefore given by
\begin{equation}\begin{split}
  \langle I\rangle(\mu,g,T) \propto\,\, &\text{Im}\{ 2 \sin(\pi g)\, K_0(T) \}\\
   &= (2 \pi T)^{2 g-1}\, B(g - i \beta \mu,g + i \beta \mu)\, \sinh \pi \beta \mu\,,
\end{split}\end{equation}
where we have introduced $\mu = \tilde e U$. In the limit $\mu\gg T$, we find that
\begin{equation}
  \langle I\rangle(\mu,g,\mu\gg T) = \text{sgn}(\mu) |2\pi \mu|^{2g-1} \frac{\pi}{\Gamma(2g)}\, .
\end{equation}
Hence, for large voltage drops, we obtain Ohms law only in the case of {\it free Fermions} where $g=1$. The resistance behaves like $R \propto |\mu|^{2g-2}$.

In the limit $\mu\ll T$, we have that
\begin{equation}
  \langle I\rangle(\mu,g,\mu\ll T) = (2\pi T)^{2g-1}\, \frac{\pi \mu}{T}\, B(g,g)\, .
\end{equation}
Thus, Ohms law is valid for all $g$ in that case. For small voltages, the resistance behaves like $R \propto T^{2g-2}$.

\subsection{Interference integral}\label{app:interferenceintegral}

For the interference current, we need to calculate the integral
\begin{equation}\begin{split}\label{eq:K1}
  K_1(x) &= (\pi T)^{2g} \int_0^{\infty} \frac{ ds \, e^{2 \pi i \mu s }}{[ \sinh (\pi T s) \sinh \pi T (s+x)]^{g}}\\
   &= (2 \pi T)^{2 g} e^{-\pi g T x} \int_0^\infty \frac{ds\, e^{2 \pi s (i \mu - g T)}}{[1 - e^{-2 \pi T s}]^{g}[1 - e^{-2 \pi T (s+x)} ]^{g}}\, .
\end{split}\end{equation}
Substituting $u = e^{- 2 \pi T s}$, we have
\begin{equation}
  K_1(x) = (2 \pi T)^{2g-1} e^{-\pi g T x} \int_0^{1} du\, u^{g - i\beta\mu - 1}\, (1-u)^{-g} (1 - e^{-2 \pi T x} u )^{-g}\,.
\end{equation}
This can be recognized as Gauss hypergeometric function, which has the integral representation \cite{WhittakerWatson}
\begin{equation}\label{eq:2F1}
  _2 F_1(a, b, c; z) = \frac{1}{B(a,c-a)}\, \int_0^1 du\, u^{a-1} (1-u)^{c-a-1} (1 - z u)^{-b}\, .
\end{equation}
Hence, we find that
\begin{equation}\label{eq:K11}\begin{split}
  K_1(x) =\,\, &(2 \pi T)^{2g-1} e^{-\pi g T x}\,\\ &B(g - i\beta\mu, 1 - g)\, _2 F_1(g - i\beta \mu, g; 1 - i\beta \mu; e^{-2\pi T x})\, .
\end{split}\end{equation}
As $z\rightarrow 1$ in \eqref{eq:2F1},
\begin{equation}
  _2 F_1(a,b;c;z = 1) = \frac{B(a,c-a-b)}{B(a,c-a)}\, .
\end{equation}
Therefore, in the limit $x\rightarrow 0$, \eqref{eq:K11} reduces to \eqref{eq:K000}, as it should; [the integral \eqref{eq:K1} reduces to the integral \eqref{eq:K0m}]. Finally, using the formulae for the Beta function above, we find
\begin{equation}\begin{split}\label{eq:K111}
  K_1(x) =
  (2 \pi T)^{2g-1} e^{-\pi g T x} \frac{\pi}{\sin\pi g} \frac{\Gamma(g-i\beta\mu)}{\Gamma(g)}\, \frac{_2 F_1(g - i\beta\mu, g; 1-i\beta\mu; e^{-2\pi T x})}{\Gamma(1-i\beta\mu)}\, .
\end{split}\end{equation}

The low-temperature limit of the interference integral can be computed in a similar way:
\begin{equation}\begin{split}\label{eq:K1low}
  K_1(x, T\ll \{\mu, x^{-1}\}) = \int_0^{\infty} ds \, [ s (s+x)]^{-g} e^{2 \pi i \mu s }\, .
\end{split}\end{equation}
Upon changing variables, $s = \frac{x}{2}(u-1)$, we obtain
\begin{equation}\begin{split}
  K_1(x, T\ll \{\mu, x^{-1}\}) = \left(\frac{2}{x}\right)^{2g-1} e^{-i\pi\mu x} \int_1^{\infty} \frac{du}{(u^2-1)^{g}}\, e^{i \pi \mu x u }\, .
\end{split}\end{equation}
As before, the limit $T\gg\mu$ can easily be found by setting $\beta\mu=0$ in \eqref{eq:K111}.

Let us now compute the interference current. From \eqref{eq:currentintegral3},
\begin{equation}\begin{split}
  \langle I \rangle &= 2 \sin(\pi g)\, (\pi T)^{2 g}\\
   &\text{Im} \{ e^{2\pi i \Phi} ( -\int_{-\infty}^{\tau^1} ds + \int_{\tau^2}^\infty ds )\, e^{2\pi i \mu s}\, [\sinh \pi T (s-\tau^1) \sinh \pi T (s-\tau^2)]^{-g} \} \, .
\end{split}\end{equation}
We have made a labeling of the edge such that $\tau^1\leq\tau^2$. Upon an appropriate shift of integration variables, the current is brought to the form \eqref{eq:K1},
\begin{equation}\begin{split}
  \langle I \rangle &= \text{Im} \{ ( e^{2 \pi i (\Phi + \mu\tau^2)} + e^{- 2 \pi i (\Phi + \mu\tau^1)})\, 2 \sin(\pi g)\, K_1(\tau^2 - \tau^1) \}\\
  &= 2 \cos 2\pi(\Phi + \frac{\mu}{2}[\tau^1+\tau^2] )\, \text{Im} \{ e^{i\pi\mu(\tau^2 - \tau^1)} \, 2 \sin(\pi g)\, K_1(\tau^2-\tau^1) \}\,,
\end{split}\end{equation}
which is the result \eqref{eq:currentintegral4}.

In the low-temperature limit, using \eqref{eq:K1low}, we obtain
\begin{equation}
  \langle I \rangle = 2 \cos 2\pi(\Phi + \frac{\mu}{2}[\tau^1+\tau^2])\, 2 \sin(\pi g)\, \int_1^{\infty} \frac{du}{(u^2-1)^{g}}\, \sin ( \pi \mu [\tau^2-\tau^1] u)\,.
\end{equation}
This integral is a Bessel function since (\cite{AbramowitzStegun}, p.~360)
\begin{equation}
  \int_1^{\infty} \frac{du\, \sin( z u)}{(u^2-1)^{g}} = \frac{\sqrt{\pi}\, \Gamma(1-g)}{2}\left(\frac{2}{z}\right)^{\frac{1}{2}-g}\, J_{g-\frac{1}{2}}(z)\,,
\end{equation}
which brings us to the result \eqref{eq:currentintegral5}.

\subsection{Interference integral with two velocities\label{app:2velocities}}

For the interference current with two velocities, we must evaluate the following integrals.
\begin{equation}
  K_2(x,y) = (\pi T)^{2g}\int_0^\infty \frac{ds\, e^{2\pi i \mu s}}{[\sinh(\pi T s)\sinh\pi T (s+x)]^{g_1} [\sinh\pi T(s+y)]^{2 g_2}}\,,
\end{equation}
for $x>0$, $y>0$. Substituting $u=e^{-2\pi T s}$, one gets
\begin{equation}\begin{split}
  K_2(x,y) = &(2 \pi T)^{2g-1} e^{-\pi T(g_1 x + 2g_2 y)}\\&\int_0^1 \frac{du\, u^{g-i\mu\beta-1}}{[(1-u)(1-e^{-2\pi T x} u)]^{g_1} [1 - e^{-2\pi T y} u]^{ 2 g_2}}\,,
\end{split}\end{equation}
which can be written in terms of Appell's hypergeometric function, $F_1$; \cite{WhittakerWatson}. We find that
\begin{equation}\label{eq:K2eval}
  K_2(x,y) = (2 \pi T)^{2g-1} e^{-\pi T(g_1 x + 2g_2 y)} \tilde F_1(g - i\mu\beta; g_1, 2 g_2; 1+g_2-i\mu\beta; e^{-2\pi T x}, e^{-2\pi T y})\,,
\end{equation}
where
\begin{equation}\label{eq:regF1}
  \tilde F_1(a;b_1,b_2;c; z_1, z_2) = B(a,c-a) F_1(a;b_1,b_2;c; z_1, z_2)\,,
\end{equation}
and $B(a,c)$ is the Euler beta function.

We also need the integral
\begin{equation}
  \hat K_2(x,y) = (\pi T)^{2g}\int_0^x \frac{ds\, e^{2\pi i \mu s}}{[\sinh \pi T (x-s)\sinh\pi T |y-s|]^{g_1} [\sinh(\pi T s)]^{2 g_2}}
\end{equation}
for $x>0$ and $y\notin (0,x)$. Substituting $u=\frac{\exp( 2\pi T s) - 1}{\exp( 2\pi T x) - 1}$, we get
\begin{equation}\begin{split}\label{eq:K2hateval}
  \hat K_2(x,y) =\, &(2\pi T)^{2g-1} \frac{ e^{\pi T g_1 (x + y)}}{[e^{2\pi T x}-1]^{g+g2-1}|e^{2\pi T y}-1|^{g_1}}\\
   &\tilde F_1(1-2g_2; g_1, 1 - g - i\mu\beta; 2-g_1-2g_2 ; \frac{e^{2\pi T x}-1}{e^{2\pi T y}-1}, 1-e^{2\pi T x})\, .
\end{split}\end{equation}

Let us now evaluate the current, with the general expression \eqref{eq:currentintegral2b}, in the case of two (families of) modes that propagate at significantly different speeds. We denote the two speeds by $u_1, u_2$ and let
\begin{equation}\label{eq:speeds}
  u_1 \ll u_2\, .
\end{equation}
Recall that $\tau^{(k)}_n = \xi^{(k)}/u_n$, where $\xi^{(1),(2)}$ is the distance between the tunneling constrictions along edge $(1)$ and $(2)$, respectively. By \eqref{eq:speeds}, $|\tau^{(k)}_1| \gg |\tau^{(k)}_2|$. In the following, we evaluate the integral in the expression for the current \eqref{eq:currentintegral2b} in the limit $\tau^{(k)}_2\rightarrow 0$.

For a Fabry-Perot interferometer, $\tau^{(1)}_n\tau^{(2)}_n < 0$. Therefore we can choose $\tau^{(1)}_1 \ll \tau^{(1)}_2 \lesssim 0 \lesssim \tau^{(2)}_2 \ll \tau^{(2)}_1$, and the integral \eqref{eq:currentintegral2b} becomes
\begin{equation}\begin{split}
  \langle I \rangle &= (\pi T)^{2g}\text{Im}\, e^{2\pi i \Phi} \{ \sin(\pi g) (\int_{-\infty}^{\tau^1_1} ds - \int_{\tau^2_1}^\infty ds ) - \sin(\pi g_2) \int_{\tau^1_1}^{\tau^2_1} f(\tau^1_2,\tau^2_2,s) ds\\
   & e^{2\pi i \mu s}\, [\sinh \pi T |s-\tau^1_1| \sinh \pi T |s-\tau^2_1|]^{-g_1} [\sinh \pi T |s-\tau^1_2| \sinh \pi T |s-\tau^2_2|]^{-g_2} \} \, .
\end{split}\end{equation}

Next, we consider the limit $\tau^{(k)}_2\rightarrow 0$. One finds that
\begin{equation}\begin{split}
  \langle I \rangle &= (\pi T)^{2g}\text{Im}\, e^{2\pi i \Phi} \{ \sin(\pi g) (\int_{-\infty}^{\tau^1_1} ds - \int_{\tau^2_1}^\infty ds ) - \sin(\pi g_2) \int_{\tau^1_1}^{\tau^2_1} \text{sgn}(s)\, ds\\
   & e^{2\pi i \mu s}\, [\sinh \pi T |s-\tau^1_1| \sinh \pi T |s-\tau^2_1|]^{-g_1} [\sinh (\pi T s)]^{-2 g_2} \} \, .
\end{split}\end{equation}
Using the $K_2$-integrals, this current is given by
\begin{equation}\begin{split}
  \langle I \rangle = \text{Im}\,\{ &e^{2\pi i \Phi} [ \sin(\pi g)\, e^{2\pi i \mu \tau^2_1} K_2(\tau^2_1-\tau^1_1,\tau^2_1) + \sin(\pi g_2) \hat K_2(\tau^2_1,\tau^1_1)]\\
   + &e^{-2\pi i \Phi} [ \sin(\pi g)\, e^{-2\pi i \mu \tau^1_1} K_2(\tau^2_1-\tau^1_1,-\tau^1_1) + \sin(\pi g_2) \hat K_2(-\tau^1_1, -\tau^2_1)]\}\, .
\end{split}\end{equation}

For a symmetric FP interferometer with $\tau^2_1 = -\tau^1_1 = \tau$, our result is
\begin{equation}\begin{split}\label{eq:IFPsym}
  \langle I \rangle = 2\cos(2\pi \Phi) \text{Im}\,\{  \sin(\pi g)\, e^{2\pi i \mu \tau} K_2(2 \tau ,\tau) + \sin(\pi g_2) \hat K_2(\tau,-\tau) \}\,.
\end{split}\end{equation}
For a strongly asymmetric FP interferometer with $|\tau^1_1| \ll \tau^2_1 = \tau$, the interference current is
\begin{equation}\begin{split}\label{eq:IFPasym}
  \langle I \rangle = \text{Im}\,\{ &e^{2\pi i \Phi} [ \sin(\pi g)\, e^{2\pi i \mu \tau} K_2(\tau,\tau) + \sin(\pi g_2) \hat K_2(\tau,0)]\\
   + &e^{-2\pi i \Phi} \sin(\pi g)\, K_2(\tau,0)\}\, .
\end{split}\end{equation}
In the asymmetric limit, the $K_2$ integrals become ordinary hypergeometric functions,
\begin{align}
  K_2(\tau,\tau) & = (2\pi T)^{2g-1} e^{-\pi T \tau( g_1 + 2 g_2)}\,\, _2 \tilde F_1(g-i\mu\beta;g_1+2 g_2;1+g_2-i\mu\beta; e^{-2\pi T\tau})\,,\label{eq:k2tt}\\
  \hat K_2(\tau,0) & = \frac{(2\pi T)^{2g-1} e^{2\pi T \tau g_1}}{[e^{2\pi T \tau}-1]^{2g-1}}\,\, _2 \tilde F_1(1-g_1-2g_2;1-g-i\mu\beta; 2-2g; 1 - e^{2\pi T\tau})\,, \label{eq:khat2t0}\\
  K_2(\tau,0) & = (2\pi T)^{2g-1} e^{-\pi T \tau g_1 }\,\, _2 \tilde F_1(g-i\mu\beta; g_1; 1-g_2-i\mu\beta; e^{-2\pi T\tau})\, .
\end{align}
Here, $\, _2 \tilde F_1$ is the regularized version of the hypergeometric function, analog to \eqref{eq:regF1}.

In the high-temperature limit, all the $K_2$ integrals are suppressed exponentially with a factor $e^{-T\tau}$. In the limit of low temperatures, $T\ll\mu,\tau^{-1}$, the interference current is expressed in terms of the confluent forms of the hypergeometric functions. For the symmetric FP interferometer, the functions from the ``Humbert series'' are needed:
\begin{equation}
  \lim_{T\rightarrow 0} \hat K_2(\tau,-\tau) = \tau^{1-2g} B(\cdot) \Phi_1(\cdot)
\end{equation}
where $\Phi_1$ is the confluent hypergeometric function of two variables. The other integral is
\begin{equation}
  \lim_{T\rightarrow 0}  K_2(2\tau,\tau) = \tau^{1-2g} \int_0^\infty du\, e^{2\pi i \mu \tau u } u^{-g1} [u+1]^{-2g2} [u+2]^{-g1}\, .
\end{equation}
It should be possible to write these low-temperature limits in terms of $\Phi_1$ or a combination of other functions in the Humbert series, but we don't know how. Therefore, for the symmetric interferometer, the finite-temperature expression \eqref{eq:IFPsym} may be used to numerically compute the low-temperature limit. For the strongly asymmetric interferometer, the low-temperature limits are
\begin{align}
  \lim_{T\rightarrow 0} K_2(\tau,\tau) & = \tau^{1-2g}\, \tilde U(1-g_1,2-2g,-2\pi i \mu \tau) \,, \\
  \lim_{T\rightarrow 0} \hat K_2(\tau,0) & = \tau^{1-2g}\, \tilde M(1-g_1-2g_2,2-2g,2\pi i \mu \tau) \,, \\
  \lim_{T\rightarrow 0} K_2(\tau,0) & = \tau^{1-2g}\, \tilde U(1-g_1-2g_2,2-2g,-2\pi i \mu \tau) \,.
\end{align}
Here, $\tilde U$ and $\tilde M$ are related to Kummer's confluent hypergeometric functions $U$ (Tricomi function) and $M$ (also denoted by $_1 F_1$). Their definition and corresponding integral representations are given by
\begin{equation}\label{eq:Mtilde}\begin{split}
  \tilde M(a,c,z) = \,_1\tilde F_1(a,c,z) &= B(a,c-a) M(a,c,z)\\ &= \int_0^1 du\, u^{a-1} [u-1]^{c-a-1} e^{2\pi i u z}
\end{split}\end{equation}
and
\begin{equation}\label{eq:Utilde}
  \tilde U(a,c,z) = \Gamma(a) U(a,c,z) = \int_0^\infty du\, u^{a-1} [1+u]^{c-a-1} e^{-2\pi i u z}\, .
\end{equation}

\newpage
%\section*{References}

\end{document}